\documentclass[%
reprint,
superscriptaddress,
preprintnumbers,
nofootinbib,
 amsmath,amssymb,
aps,
prd,
floatfix,
]{revtex4-1} 

\usepackage{amssymb,latexsym,amsfonts,amsmath,amsthm,bm,bbold,tensor, mathtools}
\usepackage{hyperref}

\numberwithin{equation}{section}

\begin{document}

\preprint{FERMILAB-PUB-25-0712-T}

\title{Studying \texorpdfstring{$\textrm{QED}_3$}{QED3} with radial quantization on the lattice: Free limit}

\author{Peter~A.~Boyle}%
\affiliation{%
Brookhaven National Laboratory, Upton, NY 11973, USA
}%
\author{Richard~C.~Brower}%
\affiliation{%
Boston University, Boston, MA 02215, USA
}%
\author{George~T.~Fleming}%
\affiliation{%
Fermi National Accelerator Laboratory, Batavia, Illinois, 60510, USA
}%
\author{Emanuel~Katz}%
\affiliation{%
Boston University, Boston, MA 02215, USA
}%
\author{Nobuyuki~Matsumoto}%
\email{nmatsum@bu.edu}
\affiliation{%
Boston University, Boston, MA 02215, USA
}%
\author{Rohan~Misra}%
\affiliation{%
Boston University, Boston, MA 02215, USA
}%

\date{\today}

\begin{abstract}
To investigate the three-dimensional quantum electrodynamics in the radial quantization on the lattice, the lattice action is constructed and the free limit is studied on $S^2 \times \mathbb{R}$.
With the overlap fermion, it is numerically verified that the important symmetries of the theory can be realized on the lattice.
The analytic correlators are derived and compared to the lattice results, which agree including the overall normalization.
The $O(a^2)$-scaling is confirmed toward the analytic value in the continuum limit, and the number of reproduced excited states is estimated heuristically for the first few refinement levels.
Our study helps us identify the features of the theory that we can study on the icosahedral lattice without fine-tuning.
\end{abstract}

\maketitle

\section{Introduction}
\label{sec:intro}

Conformal field theory (CFT) and lattice field theory have been successful in deepening the non-perturbative understanding of quantum field theory.
Indeed, many analytic results are derived in the 2D CFT \cite{Belavin:1984vu}, while the lattice study today gives predictions of quantum chromodynamics (QCD) with competitive precision to experiments \cite{Gross:2022hyw}.
To further improve the theoretical and phenomenological understanding of non-perturbative physics, the Quantum Finite Element project \cite{Brower:2016moq} aims to study CFTs in higher dimensions from the lattice. 
To cure the notorious finite-volume effects, we employ the radial quantization \cite{Fubini:1972mf,Cardy:1984epx,Cardy:1985lth}, where the only wrap-around effect is in the radial (or temporal) direction that is exponentially suppressed and does not change the eigenstates of the dilation operator. 
The difficulty in higher dimensions is then formulating the lattice theory on a curved manifold. 
The former applications of the method include the 3D Ising model \cite{Brower:2012vg} and the $\phi^4$ theory \cite{Brower:2014daa,Brower:2020jqj}.

This paper takes a step toward studying the three-dimensional quantum electrodynamics ($\textrm{QED}_3$) in this framework as a prototype of conformal gauge theories, which have phenomenological applications in beyond-Standard-Model physics (see, e.g., \cite{LatticeStrongDynamics:2023bqp,Hasenfratz:2022qan, Butt:2024kxi} for recent studies). 
${\rm QED}_3$ has collected interests from various directions for its rich features depending on the number of fermion flavors $N_f$ (counted in two-component spinors).
In fact, the theory has parity anomaly for odd $N_f$ \cite{Deser:1981wh,Redlich:1983kn,Redlich:1983dv,Appelquist:1986qw}, while it holds a conformal window for even $N_f$. 
Recently, the conformal bootstrap was applied to the $N_f=4$ theory assuming the conformality \cite{Albayrak:2021xtd} (see the reference therein for a summary of window predictions).
It is further known to have an application to superconductivity \cite{DoreyMavromatos1990,DoreyMavromatos1992,Aitchison:1992ik,Aitchison:1994un,Farakos:1997qi},
and lattice calculations have been performed to study chiral symmetry breaking \cite{Dagotto:1988id,Dagotto:1989td,Alexandre:2001pa,Lee:2002id,Hands:2002dv,Hands:2004bh}, phase diagram \cite{Azcoiti:1993ey,Fiore:2005ps,Strouthos:2007stc}, $\beta$-function \cite{Raviv:2014xna}, and conformal properties \cite{Karthik:2015sgq,Karthik:2016ppr,Karthik:2019mrr,Karthik:2020shl}, all of which are closely related, where the interesting physics lies in the zero fermion mass limit. 
The radial picture helps us scrutinize these subjects further because the curvature of spacetime induces a mass in the propagators of the massless fields. 
These two features---exponentially suppressed wrap-around effects and a massive fermion propagator---make the radial quantization for CFTs highly desirable.

We in this paper thus aim to establish the free limit. 
We derive the coupling constants for the Wilson fermion and the Gaussian gauge action by extending the formulas for simplicial lattices developed in random lattice theory~\cite{Christ:1982ck,Christ:1982ci,10.1143/PTPS.86.322} with the radial direction. 
The overlap fermion can then be written straightforwardly with the constructed Wilson fermion~\cite{Karthik:2016ppr}, which preserves important symmetries of the theory as a result of the Ginsparg-Wilson relation.

To confirm that our lattice action gives the correct continuum limit, the lattice correlators are compared to the analytic continuum formulas. 
The lattice correlators can be evaluated without Monte Carlo simulations thanks to the bilinear form of the action.
As for the gauge sector, in particular, we calculate lattice correlators without gauge fixing by using the standard conjugate gradient method, but with a projection that removes the zero modes associated with the gauge symmetry.
We show that the lattice results agree with the analytic formulas including the overall normalization, which is further supported by confirming the $O(a^2)$-scaling of the lowest operator dimension in the correlators. 
Our derived analytic formula further allows us to heuristically estimate the number of excited states that are reproduced for a given refinement level.
The result suggests that the first few descendants can be reproduced on the icosahedral lattice without refinement, which affirms the expectation that a qualitative result is obtainable with the coarse triangulation of the sphere \cite{Lao:2023zis}. 
The icosahedral lattice can thus serve as an adequate first step to simulate the interacting theory on $S^2\times \mathbb{R}$ by circumventing the complications of non-perturbative fine-tuning for anisotropic lattices.

The rest of the paper is structured as follows.
In Sec.~\ref{sec:qed3}, we summarize the symmetries of QED$_3$ in the continuum $S^2 \times \mathbb{R}$.
We describe the discretization of $S^2 \times \mathbb{R}$ in Sec.~\ref{sec:lattice} and construct the lattice action in Sec.~\ref{sec:lattice_action}. 
The numerical results are given in Sec.~\ref{sec:tests}, where the lattice results are compared with the analytic formulas derived in Appendix~\ref{sec:derivation}.
Section~\ref{sec:discussion} is devoted to discussion.

\section{ QED3 in radial quantization}
\label{sec:qed3}

We consider even $N_f$ flavors of two-component Dirac fermions $\psi_f$ ($f=1,\cdots N_f$) on $S^2 \times \mathbb{R}$.
The continuum action on ${S^2 \times \mathbb{R}}$ is
\begin{align}
    S_{\rm cont}
    &\equiv
    \int dV
    \Big[
    \frac{1}{4g^2} F_{\mu\nu}F^{\mu\nu}
    +
    \sum_{f=1}^{N_f}
    \bar \psi_f
    \sigma^a e_a^\mu 
    (
    {\nabla}^S_\mu+iA_\mu)
    \psi_f
    \Big] 
    \label{eq:TheAction}\\
    &\equiv
    S_g + S_{N_f},
    \label{eq:massless_2comp}
\end{align}
where $dV$ is the volume element, $F_{\mu\nu} \equiv \partial_\mu A_\nu - \partial_\nu A_\mu$ is the field strength of the $U(1)$ gauge field $A_\mu$, $\sigma^a$ are the Pauli matrices, and $\nabla_\mu^S \equiv \partial_\mu + \omega_\mu^S$ is the covariant derivative for the spinor fields.
We use $(x^\mu)=(\theta,\phi,t)$ for the coordinates on $S^2 \times \mathbb{R}$ and $(y^\alpha)=(\theta,\phi)$ on $S^2$ (see Appendix~\ref{sec:fermion} for conventions). 
The metric on $S^2 \times \mathbb{R}$ is
\begin{align}
    &
    ds^2
    \equiv
    g_{\mu\nu}
    dx^\mu dx^\nu
    \equiv
    dt^2
    +
    d\theta^2 
    +
    \sin^2\theta d\phi^2,
\end{align}
where the radius is set to unity. 
It is convenient to take the vierbein as
\begin{align}
  e_\mu^1 \equiv \partial_\mu \theta,
  \quad
  e_\mu^2 \equiv \sin\theta \, \partial_\mu \phi,
  \quad
  e_\mu^3 \equiv \partial_\mu t.
\end{align}

The fermion action can be rewritten with four-component spinors. 
For $f=1,\cdots, N_f/2$, we define
\begin{align}
    \Psi_f 
    \equiv 
    \left(
    \begin{array}{c}
    \psi_f  \\
    \psi_{f+N_f/2}
    \end{array}
    \right),
    \quad
    \bar \Psi_f 
    \equiv 
    (
    \bar \psi_f,
    -\bar \psi_{f+N_f/2}
    ).
\end{align}
(Recall that $\Psi_f$ and $\bar \Psi_f$ are independent in the path integral.)
We then have
\begin{align}
    S_{N_f}
    =
    \sum_{f=1}^{N_f/2}
    \bar \Psi_f
    \gamma^a e_a^\mu (\nabla^{S'}_\mu
    +
    iA_\mu)
    \Psi_f,
    \label{eq:four_comp}
\end{align}
where $S'$ stands for the four-component spinor representation of the local Lorentz group.
We use the block-diagonal basis for the gamma matrices:
\begin{align}
    \gamma_a 
    &\equiv
    \left(
    \begin{array}{cc}
        \sigma_a &  \\
         & -\sigma_a
    \end{array}
    \right)
    \quad
    (a=1,2,3)
    \label{eq:def_gamma}
\end{align}
with
\begin{align}
    \gamma_4
    &\equiv
    \left(
    \begin{array}{cc}
         & \mathbb{1}_2 \\
        \mathbb{1}_2 & 
    \end{array}
    \right),
    \quad
    \gamma_5
    \equiv
    -\gamma_1
    \gamma_2
    \gamma_3
    \gamma_4
    =
    \left(
    \begin{array}{cc}
         & -i \mathbb{1}_2 \\
        i \mathbb{1}_2 & 
    \end{array}
    \right).
\end{align}
We can define chiral transformations with the hermitian generators $\gamma_4$, $\gamma_5$, and $\gamma_{4,5}\equiv i \gamma_4 \gamma_5$, which can be seen as the isospin Pauli matrices $\tau_i$: 
\begin{align}
    \gamma_4 = \mathbb{1}_2 \otimes \tau_1,
    \quad
    \gamma_5 = \mathbb{1}_2 \otimes \tau_2,
    \quad
    \gamma_{4,5} = \mathbb{1}_2 \otimes \tau_3.
\end{align}
$SU(N_f)$ flavor symmetry in the two-component language is translated in the four-component language as the combination of the $SU(N_f/2) \times SU(N_f/2)$ flavor symmetry that rotates the flavors independently for each two-component block and the $SU(2)$ chiral symmetries that mix the two-component blocks \cite{Pisarski:1984dj}.

Other important symmetries are the parity and time-reversal symmetries  \cite{Deser:1981wh, Appelquist:1986fd}:
\begin{align}
    &P:
    x=(\theta,\phi, t)
    \to
    x_P\equiv(\pi-\theta,\phi+\pi,t), 
    \label{eq:spt_parity} \\
    &T: 
    x=(\theta,\phi,t)
    \to
    x_T\equiv(\theta,\phi,-t).
\end{align}
For a vector field $A^\mu(x)$,
\begin{align}
    &P:
    A^\mu(x)
    \to
    (-1)^{\delta^\mu_\theta}
    A^\mu(x_P), 
    \label{eq:P_vector}
    \\
    &T:
    A^\mu(x)
    \to
    (-1)^{{\delta^\mu_t}}
    A^\mu(x_T),
    \label{eq:T_vector}
\end{align}
and for a local Lorentz vector field $e^a(x)$,
\begin{align}
    &P:
    e^a(x)
    \to
    (-1)^{\delta^a_1}
    e^a(x_P), 
    \label{eq:P_local_lorentz}
    \\
    &T:
    e^a(x)
    \to
    (-1)^{\delta^a_3}
    e^a(x_T).
\end{align}
As for the spinor fields, in the two-component formalism, we define the parity as
\begin{align}
    P: 
    \psi(x) 
    \to 
    \sigma_1 \psi(x_P),
    \quad
    \bar\psi(x) 
    \to 
    -\bar \psi(x_P) \sigma_1,
    \label{eq:parity_psi}
\end{align}
and the time-reversal as
\begin{align}
    T: 
    \psi(x) 
    \to 
    \sigma_3 \psi(x_T),
    \quad
    \bar\psi(x) 
    \to 
    -\bar \psi(x_T) \sigma_3.
    \label{eq:time_reversal_psi}
\end{align}
It is easy to see that the action~\eqref{eq:TheAction} is symmetric under $P$ and $T$.
The transformation laws for $\bar\psi$ are formally the same as for $\psi^\dagger \sigma_2$.

In the four-component formalism, for later convenience, we replace the transformation law for the fermion under parity with
\begin{align}
    P': 
\begin{cases}
    \Psi(x) 
    \to 
    \gamma_5 \gamma_1 \Psi(x_P)
    =
    i(\sigma_1\otimes \tau_1) \Psi(x_P),
    \\
    \bar\Psi(x) 
    \to 
    \bar \Psi(x_P) \gamma_1 \gamma_5
    =
    -i\bar{\Psi} (x_P)
    (\sigma_1\otimes \tau_1),
\end{cases}
    \label{eq:parity_four_comp}
\end{align}
and under time-reversal with
\begin{align}
    T': 
    \begin{cases}
    \Psi(x) 
    \to 
    \gamma_5 \gamma_3 \Psi(x_T)
    =
    i(\sigma_3 \otimes \tau_1)
    \Psi(x_T) ,
    \\
    \bar\Psi(x) 
    \to 
    \bar \Psi(x_T) \gamma_3 \gamma_5
    =
    -i\bar{\Psi} (x_T)
    (\sigma_3\otimes \tau_1).
    \end{cases}
\label{eq:timerev_four_comp}
\end{align}
We can see from Eq.~\eqref{eq:four_comp} that $P'$ and $T'$ are symmetries of the action.
In fact, $P'$ can be seen as $P$ and a chiral rotation in succession:
\begin{align}
    i(\sigma_1\otimes \tau_1) \Psi(x_P)
    =
    (i\mathbb{1}_2 \otimes \tau_1)
    (\sigma_1 \otimes \mathbb{1}_2)
    \Psi_f(x_P),
    \nonumber\\
    -i\bar\Psi(x_P)
    (\sigma_1\otimes \tau_1)
    =
    \bar\Psi_f(x_P)
    (-\sigma_1 \otimes \mathbb{1}_2)
    (i\mathbb{1}_2 \otimes \tau_1).
    \label{eq:rel_two_four_parity}
\end{align}
Therefore, when the chiral symmetry is broken on the lattice, $P$ and $P'$ do not hold simultaneously.
The same argument holds for $T$ and $T'$.

\section{Lattice on \texorpdfstring{$S^2 \times \mathbb{R}$}{S2xR} }
\label{sec:lattice}

\begin{figure}[h]
    \centering
    \includegraphics[width=4.4cm]{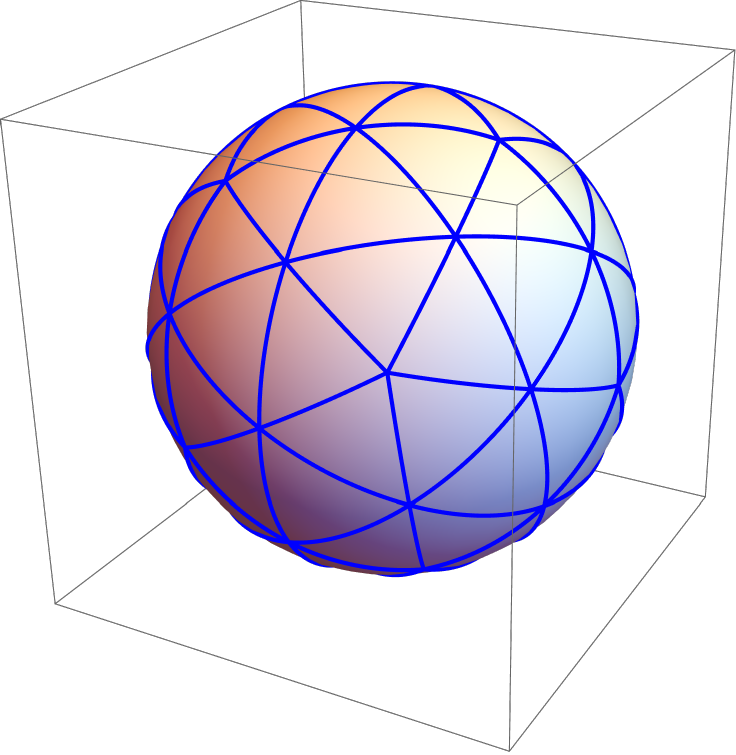}
    \caption{The tessellated $S^2$ with the refinement level $L=2$.
    }
    \label{fig:lattice_S2}
\end{figure}

In this section, we tessellate the sphere with a refined icosahedron \cite{Brower:2014daa,Brower:2016vsl} and introduce geometric quantities that are used in the later sections.
For each of the twenty triangles that make up the icosahedron, we divide the edges into $L$ segments of equal length.
By drawing lines parallel to the edges, we obtain a triangular lattice on each of the twenty original faces.
The piecewise-flat, simplicial lattice is then projected onto $S^2$, which results in the curved lattice with $N_V = 10L^2+2$ sites, $N_E = 30L^2$ links, and $N_F = 20L^2$ faces on the sphere (see Fig.~\ref{fig:lattice_S2}).
By construction, the isometry of the lattice is the full icosahedral group $I_h$, which includes parity~\eqref{eq:spt_parity}.

\begin{figure}[hbt]
    \centering
    \includegraphics[width=0.49\linewidth]{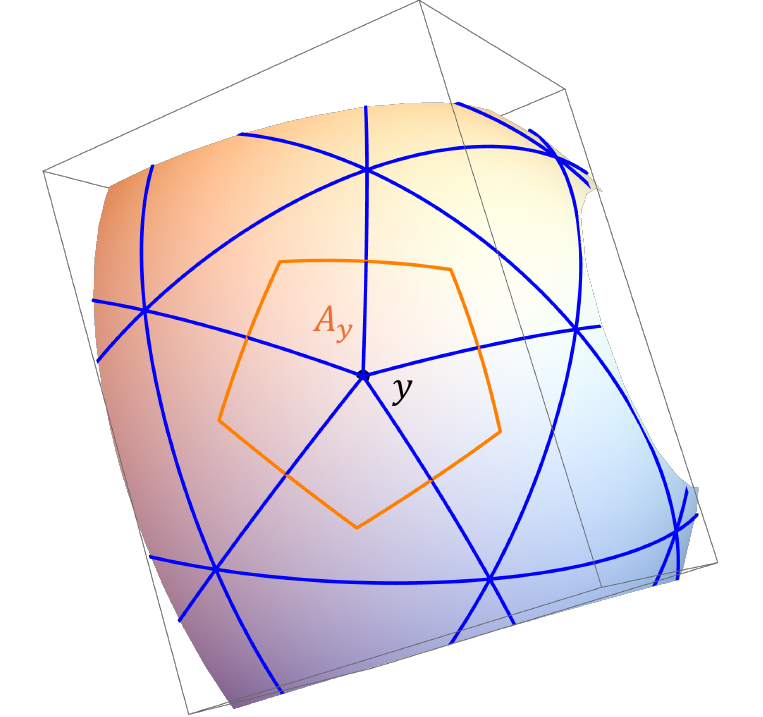}
    \includegraphics[width=0.49\linewidth]{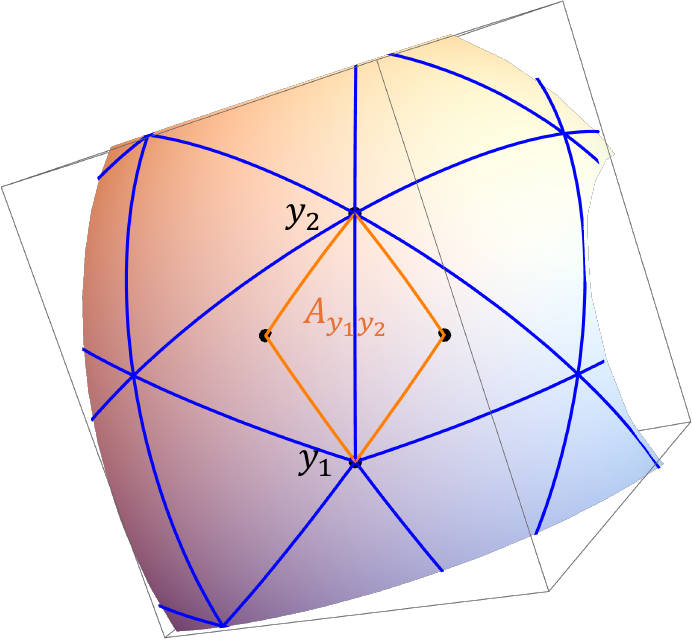}
    \caption{The area factors for (Left) the site $y$, $A_y$, and (Right) the link $y_1y_2$, $A_{y_1y_2}$.
    }
    \label{fig:areas}
\end{figure}

We write the geodesic that runs from $y_1$ to $y_2$ on $S^2$ as $\gamma_{y_1 y_2}$. 
With the affine parameter $s$ on $\gamma_{y_1 y_2}$, which increases from $y_1$ to $y_2$, we can write the directional vector of unit length as $e^\alpha_{y_1 y_2} \equiv dy^\alpha/ds$.
The lattice spin connection matrix, which is the Wilson line for the local Lorentz group, is given by
\begin{align}
    \Omega_{y_1y_2}
    = 
    \cos \frac{\omega_{y_1y_2}}{2}
    +
    i\sigma_3
    \sin\frac{\omega_{y_1y_2}}{2},
    \label{eq:expn_omega}
\end{align}
where
\begin{align}
    \omega_{y_1y_2}
    \equiv
  \int_{\gamma_{y_1 y_2}} ds \,
  e^\alpha_{y_1 y_2}
  \omega_\alpha^{12}
  =
  -\int_{\gamma_{y_1 y_2}} ds \,
  \frac{d\phi(s)}{ds}
  \cos \theta(s)
  \label{eq:expn_int_part}
\end{align}
with the spin connection $\omega_\alpha^{ab}$ calculated in the $(\theta,\phi)$ coordinates.
We can see from Eq.~\eqref{eq:expn_int_part} that the angular variable $\omega_{y_1y_2}$ changes sign under parity~\eqref{eq:spt_parity}:
\begin{align}
    \omega_{y_1^P y_2^P} = - \omega_{y_1 y_2}. 
    \label{eq:omega_parity}
\end{align}
We write the geodesic length as $\ell_{y_1y_2}$ and the average of $\ell_{y_1y_2}$ as $\bar{a}_s$, which defines the spatial cutoff scale.
The dual point of a triangular plaquette is defined by projecting the circumcenter of the triangle onto $S^2$.
We write the area factor for a site $y$ as $A_y$, which is the area of the dual plaquette that encircles $y$.
Similarly, the area factor $A_{y_1y_2}$ for a link $y_1y_2$ is the sum of the two areas of the spherical triangles that are made with the link $y_1y_2$ and the nearby dual sites (see Fig.~\ref{fig:areas}).

Note that the lengths and areas in this paper are given intrinsically on the sphere, which can be calculated with the spherical geometry formulas.
However, the effect of curvature in these quantities is $O(a^2)$ relatively, which is the order to be ignored when we construct the lattice action.
It is thus equally possible to use the flat space formulas regarding the lattice points as vertices of a simplicial lattice, which results in an action with different $O(a^2)$ coefficients.

The temporal direction can be added by simply copying the tessellated sphere $L_t$ times and
introducing the temporal lattice spacing $a_t$. 
We write as $T\equiv a_t L_t$ the physical length in the temporal direction.
The quantities that are quadratic in $\bar{a}_s$ and $a_t$ are written as $O(a^2)$ for brevity.
Note that, in the radial picture, the physical roles of the two lattice spacings $\bar{a}_s$ and $a_t$ in the system are transparent:
From the operator-state correspondence, the spatial lattice spacing $\bar{a}_s$ defines the cutoff scale for the state profile that is realized on the sphere, while the temporal lattice spacing $a_t$ can be seen as a resolution scale for the propagation of the states in the radial direction.
Note also that the dimensionful quantities are always measured with respect to the sphere radius, which we have set to unity.

\begin{figure}[htb]
  \centering
  \includegraphics[width=8cm]{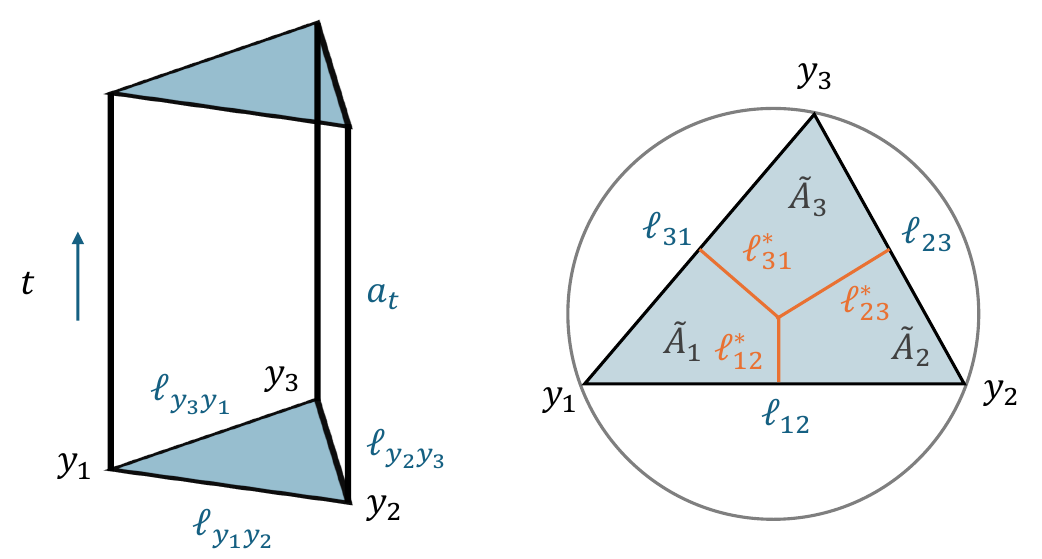}
  \caption{
   (Left) A triangular prism and (Right) the base triangle.
  The dual point of a spatial plaquette is defined by projecting the circumcenter onto the sphere.
  }
  \label{fig:prism}
\end{figure}

The fundamental volume unit on the lattice is the triangular prism as depicted in Fig.~\ref{fig:prism}.
The prism can be labeled as $(\triangle, t)$ with the base triangle $\triangle$ and the time slice label $t$.
For a base triangle that consists of the points $y_{i}$ ($i=1,2,3$), we define the dual length $\ell^*_{i,i+1}$ as the geodesic length from the dual point to the edge $y_iy_{i+1}$, where the indices $i$ are understood under modulo 3.
The length $\ell_{y_i y_{i+1}}$ is abbreviated as $\ell_{i,i+1}$.
The base triangle is divided into three pieces by the dual lines, whose areas are written as $\tilde{A}_i$.

\section{Lattice action}
\label{sec:lattice_action}

In this section, we construct free lattice actions on $S^2 \times \mathbb{R}$.
The Wilson fermion is defined in Sec.~\ref{sec:wilson}, with which the overlap fermion can be constructed straightforwardly as in Sec.~\ref{sec:overlap}.
The symmetry properties of the lattice fermions are summarized in Sec.~\ref{sec:lattice_symmetry}. 
The Gaussian gauge action for the non-compact $U(1)$ field is given in Sec.~\ref{sec:lattice_gauge}.

\subsection{Wilson fermion}
\label{sec:wilson}

The one-flavor massless Wilson fermion $S_W \equiv \bar \psi D_W \psi$ can be written as a radial extension from the $S^2$ case \cite{Brower:2016vsl}:
\begin{widetext}
\begin{align}
    S_W
    &\equiv
    \sum_t
    \sum_{y_1, y_2}
    \kappa_{y_1 y_2}    
    \bar \psi_{y_1,t}
    \Big[
    -\frac{1}{2}
    \big(
    1
    -
    e_{y_1 y_2}^a(y_1) \sigma_a
    \big)
    U_{y_1,t; y_2,t}
    \Omega_{y_1 y_2}
    \psi_{y_2,t}
    +
    \frac{1}{2}
    \psi_{y_1,t}
    \Big] \nonumber\\
    &+
    \sum_t
    \sum_y
    \kappa'_y
    \bar \psi_{y,t}
    \Big[
    - \frac{1}{2}
    (1-\sigma_3)
    U_{y,t; y,t+1}
    \psi_{y,t+1}
    - \frac{1}{2}
    (1+\sigma_3)
    U_{y,t; y,t-1}
    \psi_{y,t-1}
    +
    \psi_{y,t}
    \Big],
    \label{eq:Wilson}
\end{align}
\end{widetext}
where $\kappa_{y_1 y_2}$ are the spatial hopping parameters that are nonzero only if $(y_1, y_2)$ are nearest neighbors, $\kappa'_y$ the temporal hopping parameters, and $e_{y_1 y_2}^a \equiv e_{y_1 y_2}^\mu e_\mu^a$.
In the free limit, the couplings can be chosen as
\begin{align}
    \kappa_{y_1 y_2}
    =
    \frac{ 2A_{y_1 y_2} }
    { 
    \bar{a}_s
    \ell_{y_1 y_2} 
    },
    \quad
    {\kappa}'_y
    =
    \frac{A_{y}}{\bar{a}_s a_t}.
    \label{eq:hopping}
\end{align}
The spatial coupling is the same as the simplicial lattice formula in Ref.~\cite{Christ:1982ci}.
The temporal coupling is chosen such that we have the continuum limit with the correct spinor structure in each fundamental volume.

To see that the above choice gives the correct continuum limit, we observe that the lattice action can be rearranged into a sum over prisms $(\triangle, t)$:
\begin{align}
    S_W=
    \sum_{(\triangle, t)}
    \bar{\psi}(
    C^{(\triangle, t)}
    +
    B^{(\triangle, t)}
    )
    \psi,
    \label{eq:local_action}
\end{align}
where $\bar\psi C^{(\triangle, t)} \psi$ is the naive fermion term:
\begin{widetext}
\begin{align}
    \bar\psi C^{(\triangle, t)}
    \psi
    &\equiv
    \sum_{i=1,2,3}
    \frac{\ell^*_{i,i+1}}{2\bar{a}_s}
    \left\{
    \begin{array}{c}
    \bar \psi_{i,t}
    \,
    \Omega_{i,i+1/2}
    \,
    e_{i,i+1}^a (y_{i+1/2})
    \sigma_a 
    \,
    \Omega_{i+1/2,i+1}
    \,
    \psi_{i+1,t}\nonumber\\
    +
    \bar \psi_{i+1,t}
    \,
    \Omega_{i+1,i+1/2}
    \,
    e_{i+1,i}^a (y_{i+1/2})
    \sigma_a 
    \,
    \Omega_{i+1/2,i}
    \,
    \psi_{i,t}
    \end{array}
    \right\}
    \nonumber\\
    &+
    \sum_{i=1,2,3}
    \frac{\tilde{A}_i}{4 \bar{a}_s a_t}
    \Big\{
    \bar \psi_{i,t}
    \sigma_3
    (\psi_{i,t+1} - \psi_{i,t-1})
    -
    (
    \bar \psi_{i,t+1}
    -
    \bar \psi_{i,t-1}
    )
    \sigma_3
    \psi_{i,t}
    \Big\},
\end{align}
\end{widetext}
and the remainder $\bar\psi B^{(\triangle, t)} \psi$ is the Wilson term.
We abbreviate $\psi(y_i,t)\equiv \psi_{i,t}$ for simplicity, and $y_{i+1/2}$ is the intersection of the simplicial and dual links, which is the midpoint of $y_iy_{i+1}$ up to the $O(\bar{a}_s^2)$ curvature effect. 
As in the conventional Wilson fermion on the flat space~\cite{Wilson:1975id}, $C^{(\triangle, t)}$ can be understood as a symmetric finite difference operator and $B^{(\triangle, t)}$ a second-order finite difference operator with hermitian spinor structures.
Accordingly, $C^{(\triangle, t)}$ and $B^{(\triangle, t)}$ are antihermitian and hermitian, which result in the imaginary and real parts of the spectrum, respectively.

As usual, we can study the action of the matrix $C^{(\triangle, t)}$ in terms of differential operators to a smooth field $\psi$ that can be expanded around the center of the prism (see Ref.~\cite{Brower:2016vsl} for a further argument using a linear finite element).
We apply for each edge
\begin{align}
    \psi_{i+1,t}
    =
    \Big(
    1
    +
    \frac{1}{2}
    \ell_{i,i+1}
    e_{i, i+1}^\mu(y_{i+1/2})
    \,
    \partial_\mu
    \Big)
    \psi(
    y_{i+1/2}, t
    )
    +
    O(\bar{a}_s^2).
\end{align}
Then for the expanded coefficients, the simplicial formula~\cite{Christ:1982ck,Christ:1982ci,10.1143/PTPS.86.322} can be utilized:
\begin{align}
    \sum_{i=1,2,3}
    \ell_{i,i+1}
    \ell_{i,i+1}^*
    e^{a}_{i,i+1}
    e^{b}_{i,i+1}
    =
    A_\triangle 
    \delta^{ab}
    \cdot
    \big(
    1+
    O(\bar{a}_s^2)
    \big)
    ,
    \label{eq:triangle_formula}
\end{align}
where the relative $O(\bar{a}_s^2)$ error is the curvature effect.
We thus obtain
\begin{align}
    C^{(\triangle, t)}
    =
    \frac{1}{\bar{a}_s a_t}
    \cdot
    A_\triangle a_t
    \cdot
    \sigma^a 
    e_a^\mu
    \overleftrightarrow{\nabla}_\mu^S
    \cdot
    \big( 1 + O(a^2) \big),
    \label{eq:C_cont}
\end{align}
where $\bar\psi \overleftrightarrow{\nabla}_\mu^S \psi
\equiv
(1/2)[\bar\psi \nabla_\mu^S \psi
-
\bar\psi \overleftarrow{\nabla}_\mu^S \psi]$
and
$\bar\psi \overleftarrow{\nabla}_\mu^S \equiv \partial_\mu \bar\psi - \bar\psi \omega^S_\mu$.
The Wilson term is $O(a^2)$ and has a zero at the physical pole and is nonzero at the doubler poles.
For the regularly shaped lattices of our interest, $B^{(\triangle, t)}$ is bounded from below and removes the doublers sufficiently.

\begin{figure*}[htb]
    \centering
    \includegraphics[width=0.42\linewidth]{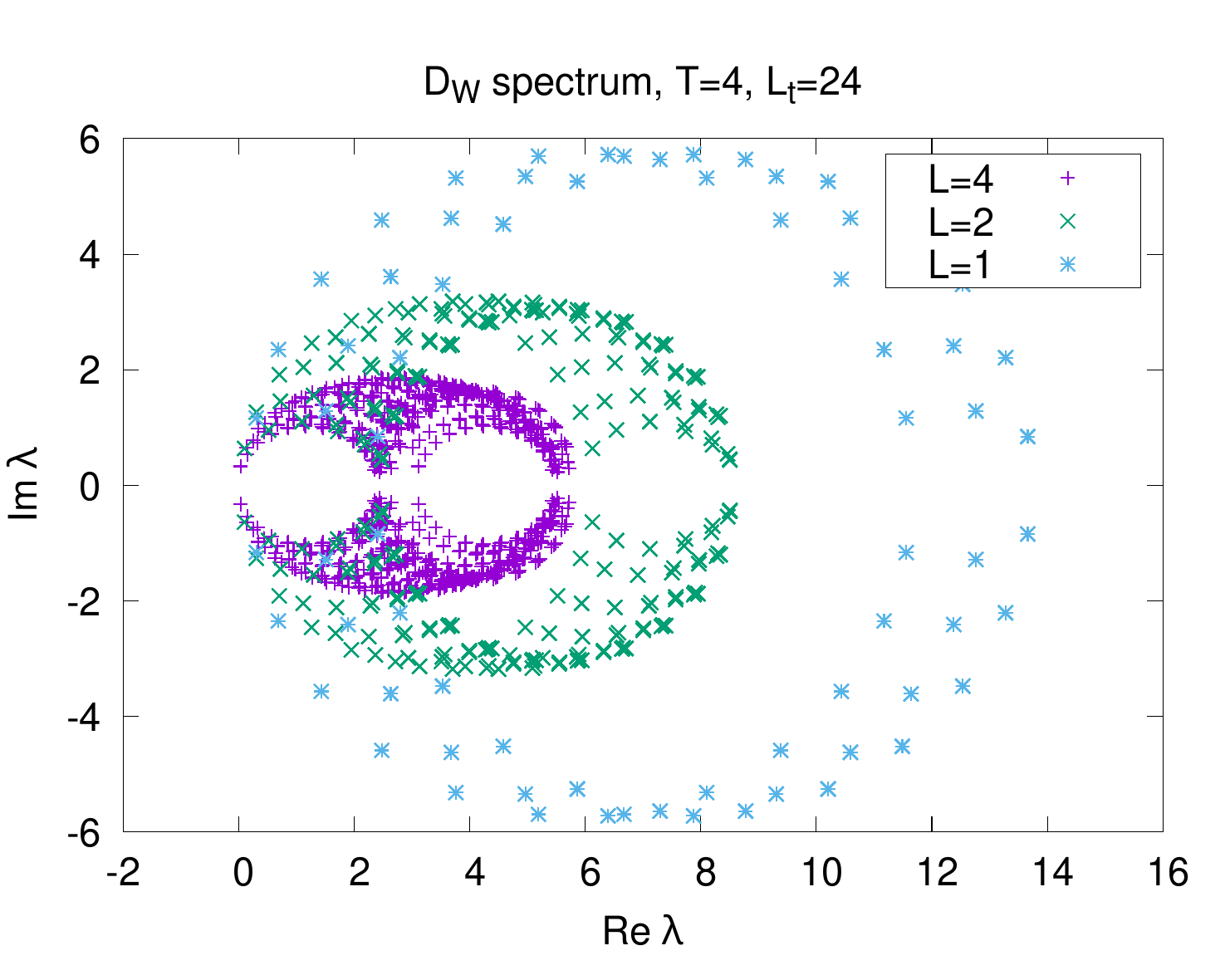}
        \hspace{0.05\linewidth}
    \includegraphics[width=0.42\linewidth]{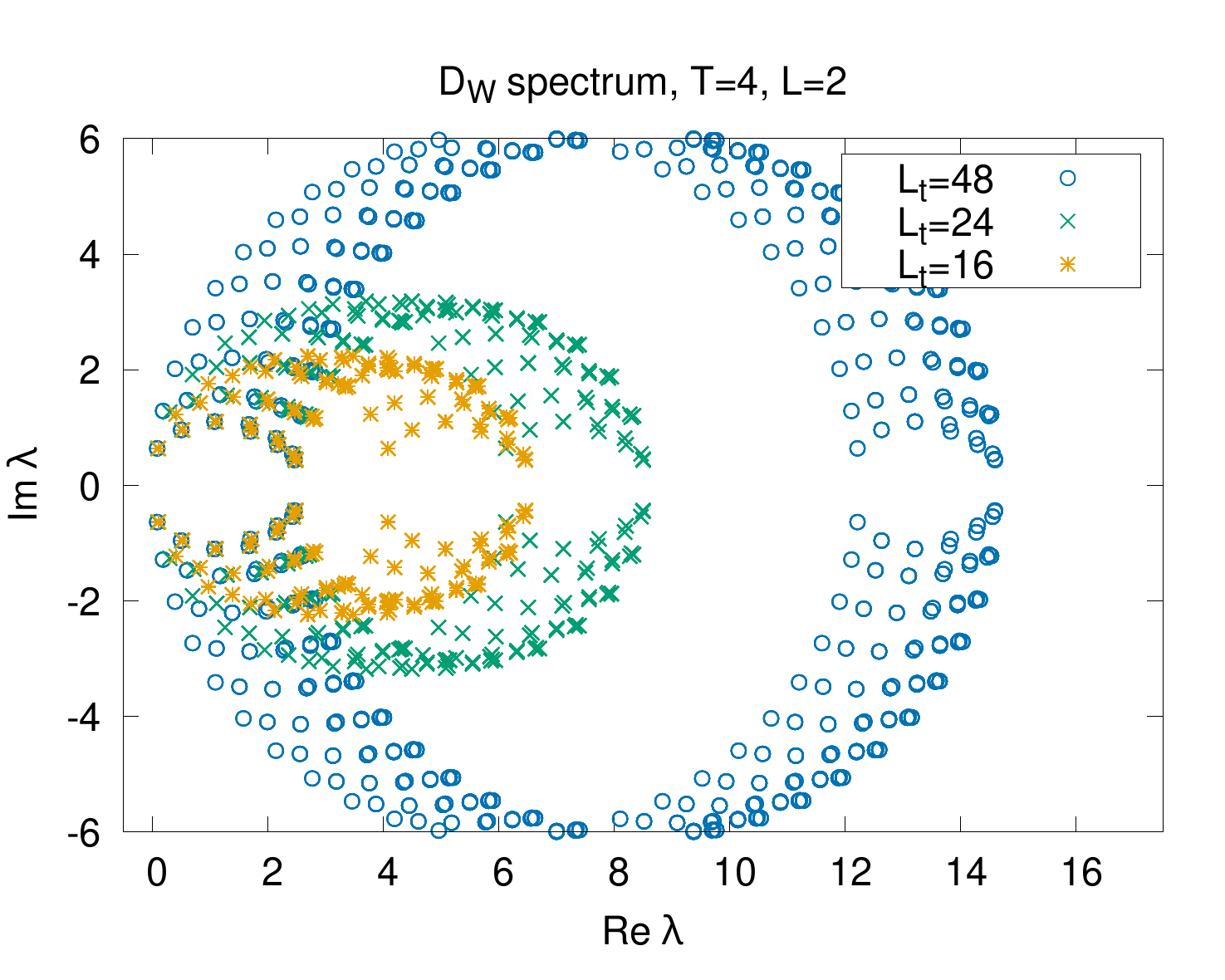}
    \caption{Spectrum of the free Wilson-Dirac operator $D_W$ with $T=4$.
    (Left) With various refinement levels $L=1,2,4$ for a fixed $L_t=24$.
    (Right) With various temporal extent $L_t=16, 24, 48$ for a fixed $L=2$.
    }
    \label{fig:eig_W}
\end{figure*}

In Fig.~\ref{fig:eig_W}, we show the spectrum of the operator $D_W$ with varying refinement levels $L$ and temporal extent $L_t$ for $T=4$. 
All the curved-lattice spectra in this paper are numerically obtained with the cuSOLVER library by treating $D_{\rm lat}$ as a dense matrix.
The hermitian and antihermitian parts of the operator give the real and imaginary parts of $\lambda$, respectively.
As in the $S^2$ case \cite{Brower:2016vsl}, the spectrum resembles that for the flat triangular lattice with radial extension.
By performing the Fourier transform and solving the characteristic equation for the resulting spinor matrix, it reads for the continuous momentum:
\begin{widetext}
\begin{align}
    \lambda_{\rm flat}
    &=
    \kappa 
    \big(
    3-\cos k_1-\cos k_2-\cos (k_1+k_2)
    \big)
    +\kappa'
    \big(
    1-\cos k_t
    \big)
    \nonumber\\
    &~~~~\pm i \sqrt{ 
    \kappa^2 
    (
    e_{12} \sin k_1 
    + 
    e_{23} \sin k_2
    + 
    (e_{12}+e_{23}) \sin (k_1+k_2)
    )^2
    +
    \kappa^{\prime 2}
    \sin^2 k_t
    },
    \label{eq:flat_spectrum}
\end{align}
\end{widetext}
where $\kappa=1/\sqrt{3}$ and $\kappa'=(\sqrt{3}/2)({\bar{a}_s}/a_t)$ with $e_{12}\equiv(1,0)$ and $e_{23}\equiv(- 1/2,\sqrt{3}/2)$.
\begin{figure*}[hbt]
    \centering
    \includegraphics[width=0.42\linewidth]{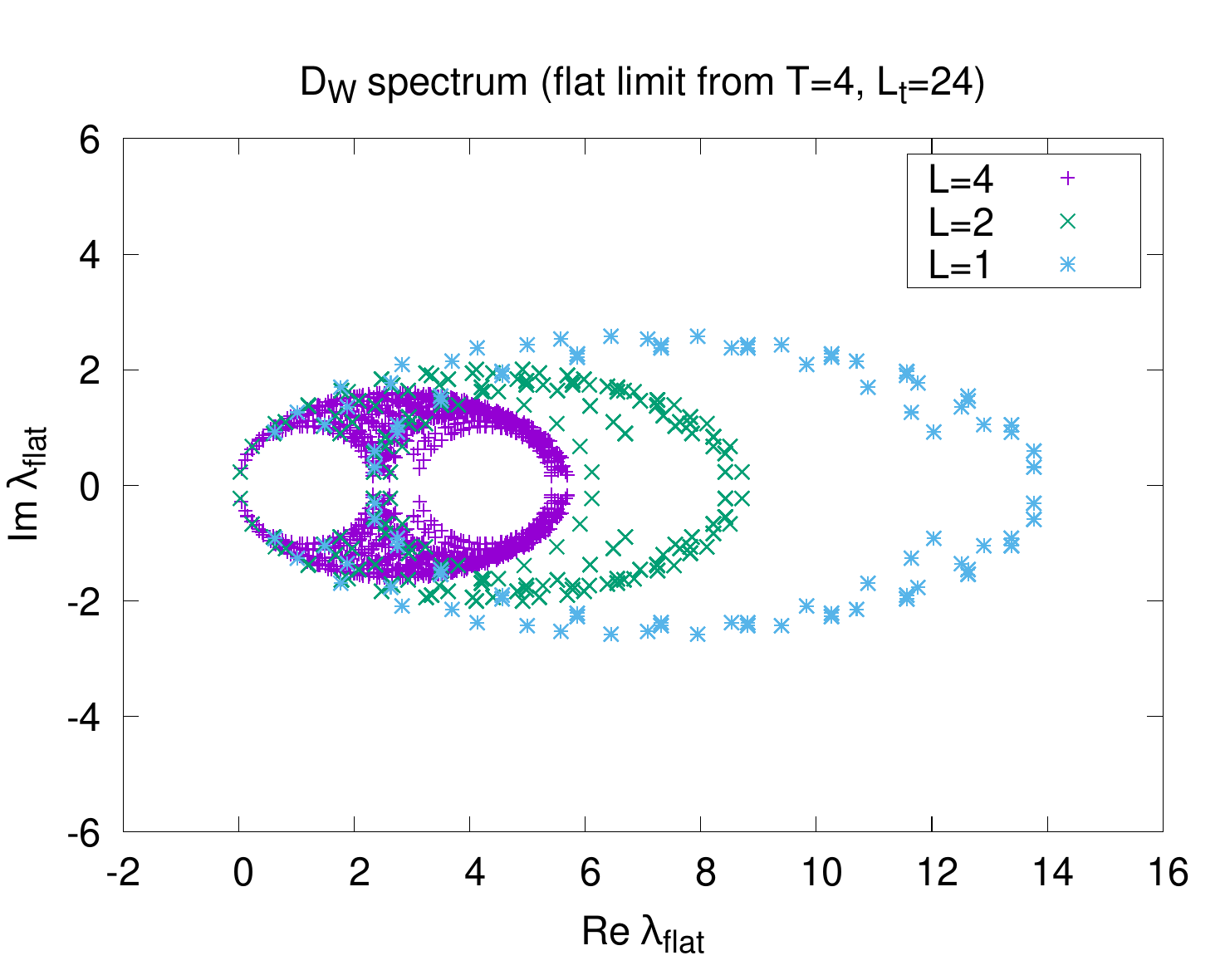}
        \hspace{0.05\linewidth}
    \includegraphics[width=0.42\linewidth]{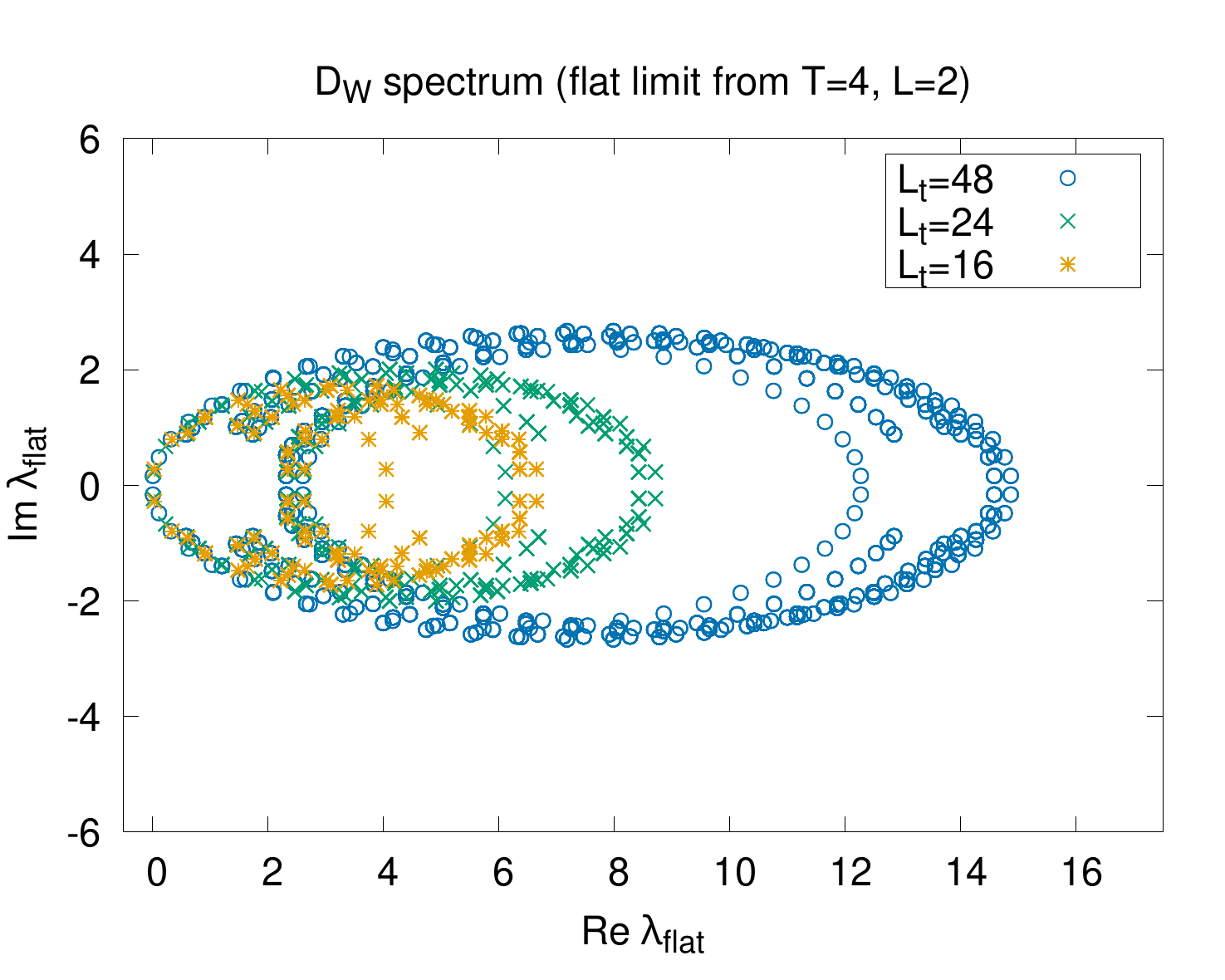}
    \caption{ The flat spectrum~\eqref{eq:flat_spectrum} of the Wilson fermion on the equilateral triangular lattice with radial extension, which we use as an approximation of the spectrum on $S^2\times\mathbb{R}$.
    The couplings are set from $\bar{a}_s$ and $a_t$ of the lattices in Fig.~\ref{fig:eig_W}.
    }
    \label{fig:flat_spectrum}
\end{figure*}
In Fig.~\ref{fig:flat_spectrum}, we plot the flat spectrum with the couplings set from the same lattice spacings $\bar a_s$ and $a_t$ for the curved lattices in Fig.~\ref{fig:eig_W}.
The number of lattice points is also adjusted to be close to that of the curved lattice.
For large $L$, where the effect of the irregular five-point vertices becomes suppressed, we see that the flat spectrum~\eqref{eq:flat_spectrum} provides a qualitative description of the curved space spectrum.
Observing that the locations of the first doubler pole agree to a good approximation between the two, which is expected because the curvature is locally an $O(\bar{a}_s^2)$ effect, we use the flat space formula ${\rm min}(4\kappa, 2\kappa') = {\rm min}(4/\sqrt{3}, \sqrt{3} {\bar{a}_s}/a_t )$ as an estimate of the location of the first doubler pole in $S^2\times \mathbb{R}$.

It is important to remark that our definition of the Wilson-Dirac operator~\eqref{eq:Wilson} fixes the location of the spatial doubler and lets the temporal doubler move according to the anisotropy $\bar{a}_s/a_t$.
Indeed, in the flat limit, the location of the temporal doubler coincides with the spatial doublers when $\bar{a}_s/a_t = 4/3$, and as $\bar{a}_s/a_t \to \infty$, the temporal doubler freezes out.

\subsection{Overlap fermion}
\label{sec:overlap}

Following Ref.~\cite{Karthik:2016ppr} (see also {Refs.~\cite{Narayanan:1994gw,Narayanan:1997by,Kikukawa:1997qh}), we define the one-flavor massless overlap fermion $S_{\rm ov} \equiv \bar \psi D_{\rm ov} \psi$ with
\begin{align}
    &D_{\rm ov}
    \equiv 
    1 + X  \frac{1}{\sqrt{X^\dagger X}},
\end{align}
where $X \equiv D_W - M$.
According to the estimate of the first doubler pole from Eq.~\eqref{eq:flat_spectrum}, we set $M$ in the range: 
\begin{align}
    0
    <
    M
    <
    \alpha M_0,
    \quad
    M_0 
    \equiv
    {\rm min}
    \Big(\frac{4}{\sqrt{3}}, 
    \frac{\sqrt{3}{\bar{a}_s}}{a_t} 
    \Big)
    \label{eq:overlapM}
\end{align}
with the safety factor, say, $\alpha=0.9$, to take into account the slightly skewed shape of the triangles.
Under the condition $\bar{a}_s/a_t \geq 4/3$, this means we can use the conventional range $0<M<2$.
In the left panel of Fig.~\ref{fig:eig_ov}, we show the spectrum of the overlap operator $D_{\rm ov}$ with $L=4$, $L_t=24$, $T=4$, and $M=1$, together with the original Wilson spectrum. 
\begin{figure*}[htb]
    \centering
    \includegraphics[width=0.42\linewidth]{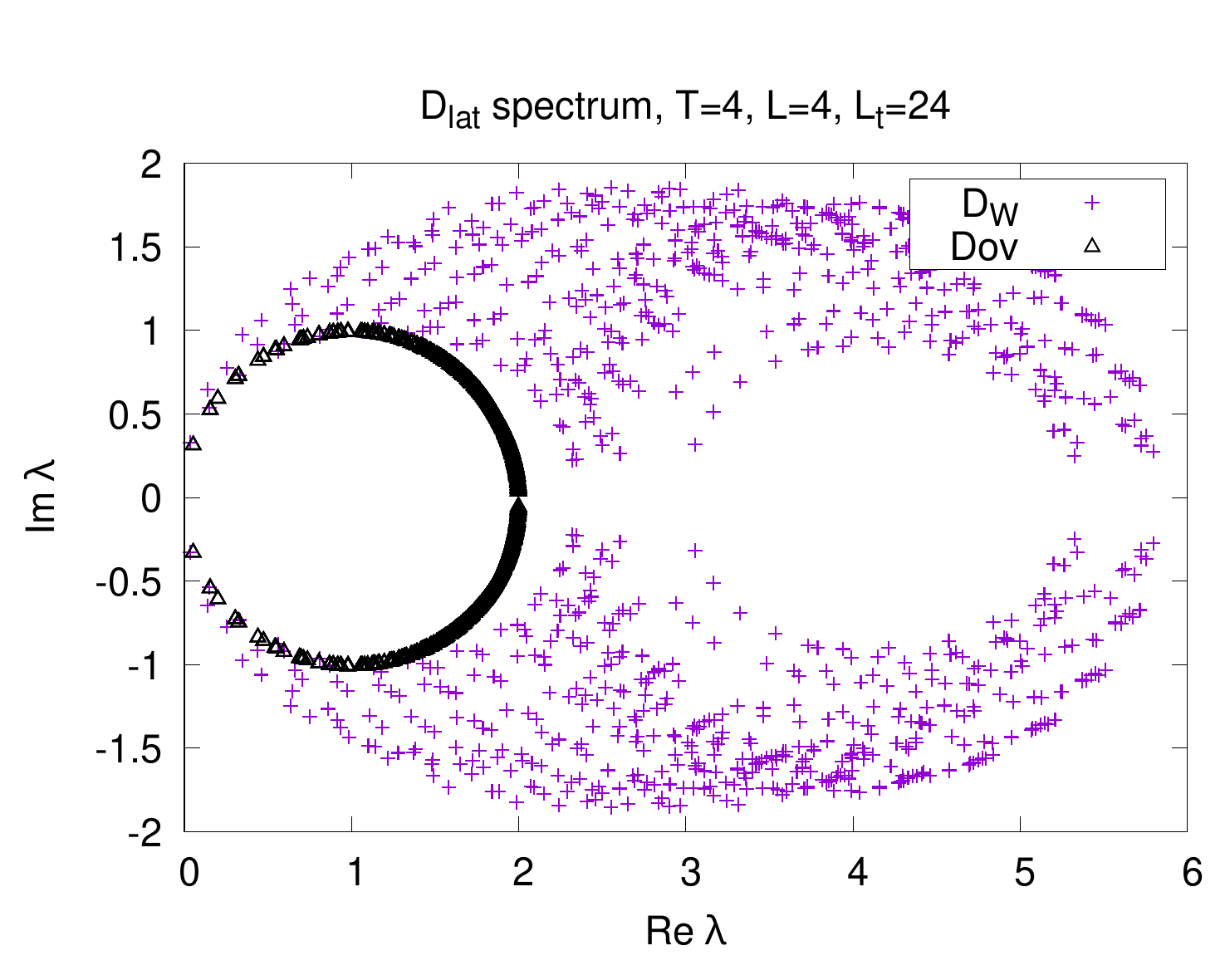}
        \hspace{0.05\linewidth}
    \includegraphics[width=0.42\linewidth]{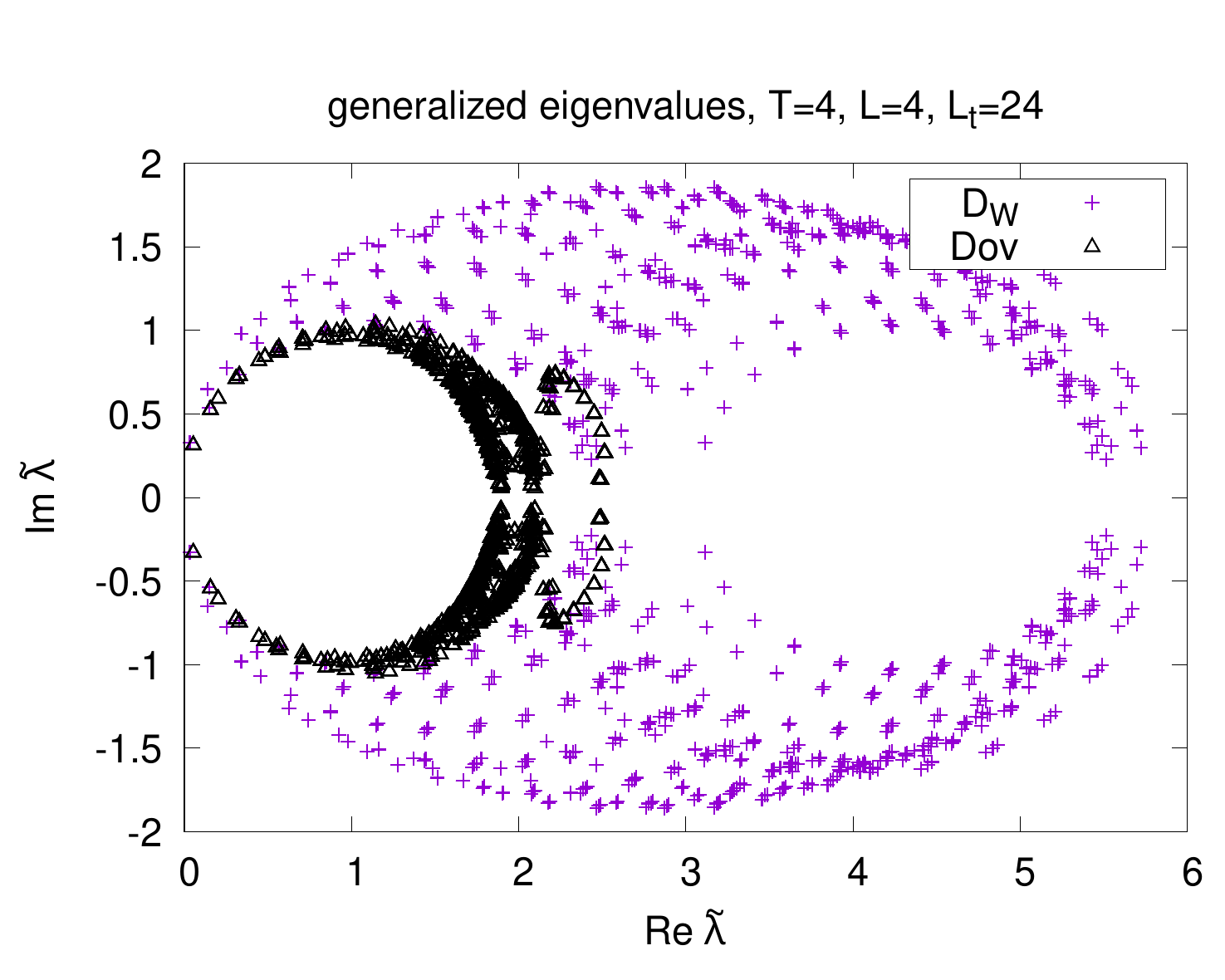}
    \caption{(Left) Comparison of the spectrum between the Wilson and overlap operators. 
    (Right) The generalized eigenvalues~\eqref{eq:gevp} that approach the continuum spectrum.
    $T=4$, $L=4$, $L_t=24$, and $M=1$ in both panels.
    }
    \label{fig:eig_ov}
\end{figure*}
We see that the Wilson spectrum is correctly projected onto the unitary circle.

It should be noted that the lattice spectrum $\lambda$ studied above differs from the continuum spectrum by a volume factor \cite{Brower:2016vsl}.
In fact, from Eq.~\eqref{eq:C_cont}, we see the correspondence:
\begin{align}
    (\bar{a}_s a_t)
    \big[ D_{\rm lat}\big]_{xx'}
    \Leftrightarrow
    \sqrt{g(x)} \, d^3x \, 
    \cdot 
    \big[ D(x) \delta^3(x-x') d^3x' \big].
\end{align}
The counterpart of the volume element $dV = \sqrt{g(x)} \, d^3x$ is given on the lattice by the volume matrix $\delta V \equiv {\rm diag}(A_y a_t)$, which is not proportional to the identity.
The effect of the varying local volume can be compensated by considering the generalized eigenvalue problem \cite{Brower:2016vsl}:
\begin{align}
    D_{\rm lat}\psi
    =
    \tilde{\lambda} \cdot ({\overline{\delta V}})^{-1}
    \delta V \psi.
    \label{eq:gevp}
\end{align}
In order to match the scaling, we divide by the average of the local volume elements, ${\overline{\delta V}}$, on the right-hand side.
The generalized eigenvalues $\tilde\lambda$ for the Wilson and overlap operators for $T=4$, $L=4$, $L_t=24$, and $M=1$ are shown in the right panel of Fig.~\ref{fig:eig_ov}.

\subsection{Global symmetries for the lattice fermions}
\label{sec:lattice_symmetry}

In this section, following Refs.~\cite{Karthik:2015sgq, Karthik:2016ppr}, we define the multi-flavor lattice action and summarize its global symmetries. 
Though we mainly focus on parity, the argument holds in parallel for time-reversal symmetry.

We recall from Sec.~\ref{sec:qed3} that the continuum Dirac operator $D(x;A) \equiv \sigma^a e_a^\mu ( {\nabla}^S_\mu+iA_\mu )$ satisfies
\begin{align}
    -\sigma_1 D(x_P;A_P) \sigma_1 = D(x;A),
    \label{eq:parity_Dcont}
\end{align}
where $A_P^\mu$ is the parity-transformed field~\eqref{eq:P_vector}, thereby making $P$ a symmetry of the action.
For simplicity, the dependence on the gauge field will be suppressed below.
The counterpart of Eq.~\eqref{eq:parity_Dcont} for the Wilson-Dirac operator $D_W$ is \cite{Karthik:2015sgq, Karthik:2016ppr}
\begin{align}
    -\sigma_1 D_W\vert_{x \to x_P} \sigma_1 = -D^\dagger_W,
    \label{eq:parity_two_wilson}
\end{align}
where we use
\begin{align}
    e^a_{y_1^P y_2^P}(y_1^P)
    =
    (-1)^{\delta^a_1}
    e_{y_1y_2}^a(y_1)
\end{align}
and [see Eq.~\eqref{eq:omega_parity}]
\begin{align}
    \Omega_{y_1^P y_2^P}
    =
    \Omega^\dagger_{y_1 y_2}
    .
\end{align}
The appearance of $-D^\dagger_W$ rather than $D_W$ on the right-hand side of Eq.~\eqref{eq:parity_two_wilson} signifies the breaking of $P$ for the Wilson fermion.

On the other hand, the breaking is minimal for the overlap fermion.
To see this, we first observe that the unitarity of $X/\sqrt{X^\dagger X}$ implies that
\begin{align}
    D_{\rm ov}^{-1}
    =
    \mathbb{1}-(D_{\rm ov}^\dagger)^{-1},
    \label{eq:ginsparg_wilson_prop}
\end{align}
where $\mathbb{1}$ is the Kronecker delta in the coordinate space and its appearance is a characteristic of the Ginsparg-Wilson fermion (see, e.g., Ref.~\cite{Kikukawa:1998pd}).
We therefore have for the overlap operator
\begin{align}
    -\sigma_1 (D_{\rm ov}\vert_{x \to x_P})^{-1} \sigma_1
    = 
    -(D^\dagger_{\rm ov})^{-1}
    =
    D_{\rm ov}^{-1} - \mathbb{1},
    \label{eq:parity_overlap_2cmp}
\end{align}
which shows that the parity symmetry holds for the overlap propagator up to the contact term.

We now include the flavor, and define the sign-definite multi-flavor action as \cite{Karthik:2015sgq, Karthik:2016ppr}
\begin{align}
    S_{N_f,{\rm lat}}
    &\equiv
    \sum_{f=1}^{N_f/2}
    \Big[
    \bar \psi_f 
    D_{\rm lat}
    \psi_f
    -
    \bar \psi_{f+N_f/2} 
    D_{\rm lat}^\dagger \psi_{f+N_f/2}
    \Big] \\
    &=
    \sum_{f=1}^{N_f/2}
    \bar \Psi_f 
    {\cal D}_{\rm lat}
    \Psi_f,
    \label{eq:lattice_multi_flavor}
\end{align}
where we have introduced the four-component lattice Dirac operator:
\begin{align}
    {\cal D}_{\rm lat}
    \equiv
    \left(
    \begin{array}{cc}
        D_{\rm lat} & \\
         & D^\dagger_{\rm lat}
    \end{array}
    \right).
    \label{eq:four_comp_dirac}
\end{align}
One can confirm that $P'$ and $T'$ symmetries in the four-component formalism, Eqs.~\eqref{eq:parity_four_comp} and \eqref{eq:timerev_four_comp}, are preserved exactly on the lattice \cite{Karthik:2015sgq, Karthik:2016ppr}.

As described in Eq.~\eqref{eq:rel_two_four_parity}, an $SU(2)$ chiral rotation relates the spacetime symmetries in the two- and four-component formalisms.
Accordingly, the breaking of the two-component parity $P$ for the Wilson fermion can be understood from the four-component perspective in terms of chiral symmetry breaking.
In fact, since $D_W^\dagger \neq -D_W$, a subgroup of the $SU(2)$ chiral rotations generated by $\gamma_4 = \mathbb{1}_2\otimes \tau_1$ and $\gamma_5 = \mathbb{1}_2\otimes \tau_2$ is broken for the four-component Wilson-Dirac operator ${\cal D}_W$.
(The $\gamma_{4,5} = \mathbb{1}_2 \otimes \tau_3$ rotation is a phase rotation with different signs for the two-component blocks, and the corresponding symmetry is preserved trivially for the diagonal operator~\eqref{eq:four_comp_dirac}.)
As a result, the $SU(N_f)$ flavor symmetry is broken to $SU(N_f/2) \times SU(N_f/2)$ for the Wilson fermion \cite{Karthik:2015sgq,Karthik:2016ppr}.

On the other hand, as already mentioned above, the preservation of the spacetime symmetries for the bulk part of the overlap propagator is a property of the Ginsparg-Wilson fermion.
Indeed, the four-component overlap operator ${\cal D}_{\rm ov}$ satisfies the Ginsparg-Wilson relation \cite{Ginsparg:1981bj} (see Ref.~\cite{Bietenholz:2000ca} for the study in odd dimensions):
\begin{align}
    \Gamma
    {\cal D}_{\rm ov}
    +
    {\cal D}_{\rm ov}
    \Gamma 
    =
    {\cal D}_{\rm ov} \Gamma {\cal D}_{\rm ov}
    \label{eq:gw}
\end{align}
for $\Gamma = \gamma_4, \gamma_5$.
As in four dimensions \cite{Luscher:1998pqa}, we can define the lattice chiral transformation:
\begin{align}
    \Psi \to e^{i\alpha\hat \Gamma} \Psi,
    \quad
    \bar\Psi \to \bar\Psi
    e^{i \alpha\Gamma},
\end{align}
where $\hat \Gamma \equiv \Gamma(1- {\cal D}_{\rm ov})$.
Together with the $SU(N_f/2) \times SU(N_f/2)$ flavor rotations and the $\gamma_{4,5}$ rotation, they generate the exact lattice $SU(N_f)$ symmetry for the overlap fermion (see Ref.~\cite{Karthik:2016ppr} for a proof in the operator formalism).

\subsection{Gauge action}
\label{sec:lattice_gauge}

A generic gauge-invariant Gaussian action
for the non-compact $U(1)$ field on the lattice is
\begin{align}
    &S_{g,{\rm lat}}
    \equiv
    \sum_{\triangle,t}
    \frac{\beta_{\triangle}}{2}
    \Big(
    \sum_{m \in p(\triangle, t)}
    \eta_m \theta_m
    \Big)^2 \nonumber\\
    &~~~~~~~~
    +
    \sum_{l,t}
    \frac{\beta_l}{2}
    \Big(
    \sum_{m \in p(l,t)} \eta_m
    \theta_m
    \Big)^2
    .
    \label{eq:lattice_gauge}
\end{align}
$p(\triangle, t)$ is the spacelike plaquette with the triangle $\triangle$ on the timeslice $t$, while $p(l,t)$ is the temporal plaquette with the link $l$ on the timeslice $t$.
The angular link variable $\theta_m$ is defined in terms of the gauge field $A_\mu$ as
\begin{align}
  \theta_m
  \equiv
  \int_{\gamma_m} dx^\mu A_\mu,
  \label{eq:theta_def}
\end{align}
where the integral is over the geodesic $\gamma_m$ that forms the link $m$.
The sign factor $\eta_m = \pm1$ in Eq.~\eqref{eq:lattice_gauge} is the relative orientation of $\gamma_m$ in the plaquette.
The group-valued variable $U_m$ in Eq.~\eqref{eq:Wilson} can be expressed with the angular variable $\theta_m$ as $U_m = \exp(i\theta_m)$.
In the free limit, the coupling constants can be chosen as
\begin{align}
    \beta_\triangle
    =
    \frac{1}{g^2}
    \frac{a_t}{A_\triangle},
    \quad
    \beta_{l}
    =
    \frac{1}{g^2}
    \frac{2A_l}{\ell_l^2 a_t}.
    \label{eq:free_coupling_gauge}
\end{align}
Again, the spatial coupling is the same as the simplicial lattice formula in Ref.~\cite{Christ:1982ci},
and the temporal coupling $\beta_l$ is given such that it reproduces the correct tensor contractions of $F_{\mu\nu}$ in the continuum limit.

To confirm the tensorial structure, note that the integral of the one-form $A_\mu dx^\mu$ around a plaquette $p$ is
\begin{align}
  \oint_{\partial p} A_\mu dx^\mu
  = 
  \frac{1}{2}
  \int_p dV \,
  \epsilon^{\mu\nu} 
  F_{\mu\nu},
  \label{eq:ointA}
\end{align}
where $\epsilon_{\mu\nu}$ is the invariant volume tensor on the two-dimensional surface $p$.
For a smooth gauge field $A_\mu$ such that the field strength $F_{\mu\nu}$ is almost constant in $p$, Eq.~\eqref{eq:ointA} can be approximated as
\begin{align}
  \oint_{\partial p} A_\mu dx^\mu
  \simeq
  \frac{1}{2}
  A_p
  \cdot
  \epsilon^{\mu\nu}
  \,
  F_{\mu \nu},
  \label{eq:approx}
\end{align}
where $A_p$ is the area of $p$.
For a spatial plaquette $p=p(\triangle,t)$,
\begin{align}
    \epsilon^{\alpha\beta}
    \epsilon^{\gamma\delta}
    =
    h^{\alpha\gamma} h^{\beta\delta}
    -
    h^{\alpha\delta} h^{\beta\gamma},
\end{align}
which leads to
\begin{align}
  \Big(
  \oint_{\partial p(\triangle,t)} A_\mu dx^\mu
  \Big)^2
  \simeq
  \frac{A_{\triangle}^2}{2}
  \cdot
  h^{\alpha\gamma}
  h^{\beta\delta}
  F_{\alpha\beta} F_{\gamma\delta}.
  \label{eq:intA_to_Fsq}
\end{align}
For a temporal plaquette $p=p(\ell_{i,i+1},t)$,
\begin{align}
    \epsilon^{\mu\nu}
    =
    e_{i,i+1}^\mu e_3^\nu 
    - e_3^\mu e_{i,i+1}^\nu,
\end{align}
giving
\begin{align}
  &\Big(
  \oint_{\partial p(\ell_{i,i+1},t)} A_\mu dx^\mu
  \Big)^2 \nonumber\\
  &~~~~~~~~\simeq
  (\ell_{i,i+1} a_t)^2
  \cdot
  e_{i,i+1}^\mu 
  e_{i,i+1}^\rho 
  e_3^\nu e_3^\sigma
  F_{\mu\nu} F_{\rho\sigma}.
  \label{eq:intA_to_Fsq2}
\end{align}
By using Eq.~\eqref{eq:triangle_formula}, we obtain for each base triangle
\begin{align}
  &\sum_{i}
  \frac{\ell_{i,i+1}^*}{\ell_{i,i+1} a_t}
  \Big(
  \oint_{\partial p(\ell_{i,i+1},t)} A_\mu dx^\mu
  \Big)^2 \nonumber\\
  &~~~~~~~~~~~~~~~~~~
  \simeq
  A_\triangle a_t
  \cdot
  h^{\mu \rho}
  e_3^\nu e_3^\sigma
  F_{\mu\nu} F_{\rho\sigma}.
\end{align}
With the $2+1$ decomposition of the metric:
\begin{align}
    g_{\mu\nu} = h_{\mu\nu} + e^3_\mu e^3_\nu,
\end{align}
we can assign the following factor for a triangular prism $(\triangle, t)$:
\begin{align}
    &\frac{A_\triangle a_t}{4}
    g^{\mu\rho}g^{\nu\sigma}
    F_{\mu\nu} F_{\rho\sigma} 
    \simeq
    \frac{1}{2}
    \Big[
    \frac{a_t}{A_\triangle}
  \Big(
  \oint_{\partial p(\triangle, t)} A_\mu dx^\mu
  \Big)^2
  \nonumber\\
    &\hspace{0.15\linewidth}
    +
  \sum_i
  \frac{\ell_{i,i+1}^*}{\ell_{i,i+1}a_t}
  \Big(
  \oint_{\partial p(\ell_{i,i+1}, t )} A_\mu dx^\mu
  \Big)^2
  \Big].
\end{align}
By summing over all prisms,
we obtain
the lattice action~\eqref{eq:lattice_gauge}
with the coupling constants~\eqref{eq:free_coupling_gauge}
under the identification~\eqref{eq:theta_def}.

\section{Numerical tests}
\label{sec:tests}

In this section, we numerically confirm the properties of the lattice action derived in Sec.~\ref{sec:lattice_action}.
For the overlap fermion, we use the Zolotarev approximation \cite{vandenEshof:2002ms,Chiu:2002eh} to regularize $X \cdot (X^\dagger X)^{-1/2}$, where the error is controlled to be on the order of machine precision. 
We set $M=1$ throughout for the overlap fermion.

\subsection{Lattice Dirac propagator}
\label{sec:dirac_propagators}

We define the continuum fermion propagator:
\begin{align}
    G(x,x') \equiv \langle \psi (x) \bar{\psi} (x') \rangle.
\end{align}
From Eq.~\eqref{eq:C_cont}, we see that the corresponding lattice propagator is
\begin{align}
    G_{\rm lat}(x, x') \equiv \frac{1}{ \bar{a}_s a_t}[D^{-1}_{\rm lat}]_{xx'}.
\end{align}
In particular, we consider the temporal propagator $G(t)$ by setting $x=(0,0,t)$ and $x'=(0,0,0)$.
In Appendix~\ref{sec:ferm_prop}, the analytic expression is derived to be
\begin{align}
    G(t)
    &=
    \sigma_3
    \cdot
    {\rm sign}(t)
    \cdot
    \frac{1}{4\pi}
    \sum_{n \geq 0} 
    (n+1)
    e^{-(n+1)\vert t \vert}.
    \label{eq:exact_Gf}
\end{align}
One can confirm explicitly the time-reversal symmetry for the propagator, where the sign function cancels the additional sign factor for $\bar\psi$ in Eq.~\eqref{eq:time_reversal_psi}.
Figure~\ref{fig:prop_ov} shows the overlap and continuum propagators, where the $(1,1)$ spinor component is taken.
\begin{figure}[htb]
    \centering
    \includegraphics[width=0.85\linewidth]{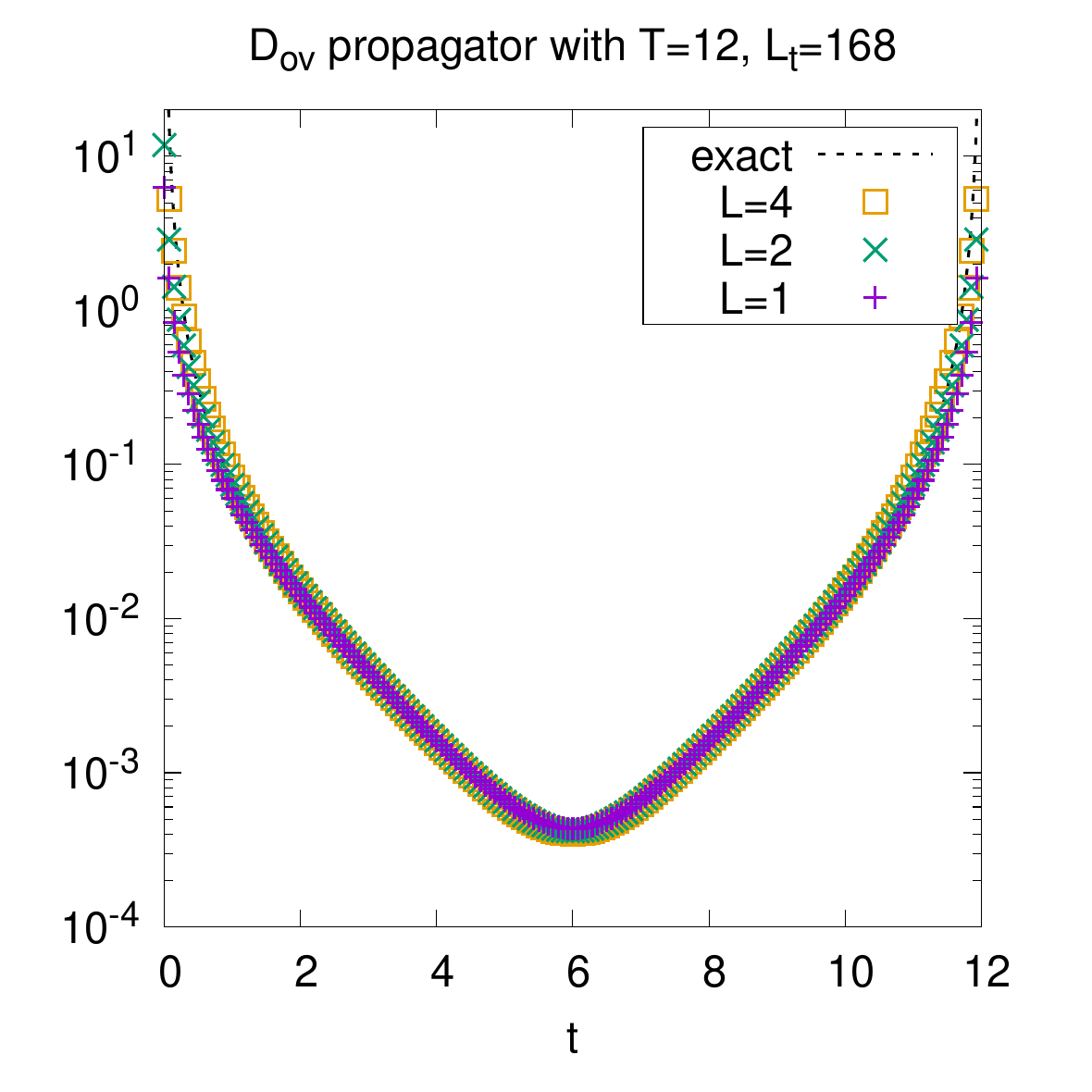}
    \caption{ The free overlap propagator with the refinement levels $L=1,2,4$ for $T=12$ and $L_t=168$.
    }
    \label{fig:prop_ov}
\end{figure}
The refinement level is varied as $L=1,2,4$ for fixed $T=12$ and $L_t=168$.
Note that the antiperiodic boundary condition cancels the minus sign in the negative $t$ direction, making the functional shape symmetric.
We confirm that the overlap propagator reproduces the continuum functional form with the correct normalization. 

We further scrutinize the continuum limit by calculating the scaling of the lowest operator dimension that contributes to the propagator.
We conventionally define the effective mass (or properly, in our context, the effective {\it dimension}) $\Delta_{\rm eff}(t)$ as
\begin{align}
    &f(t)\equiv\cosh^{-1} \Big(\frac{G^{(1,1)}_{f,\rm{ov}}(t)}{G^{(1,1)}_{f,\rm{ov}}(T/2)}\Big),
    \\
    &\Delta_{\rm eff}(t)
    \equiv
    -
    \frac{1}{a_t}\big(
    f(t+a_t)-f(t)
    \big).
    \label{eq:def_eff_mass}
\end{align}
We fit the approach of $\Delta_{\rm eff}(t)$ to the plateau with the ansatz:
\begin{align}
    \Delta_{\rm eff}(t) \simeq \Delta_0 + c e^{- \Delta't},
\end{align}
where $\Delta_0$ is the desired lowest dimension of the operator (see Fig.~\ref{fig:meff_fit}).
\begin{figure}[htb]
    \centering
    \includegraphics[width=0.8\linewidth]{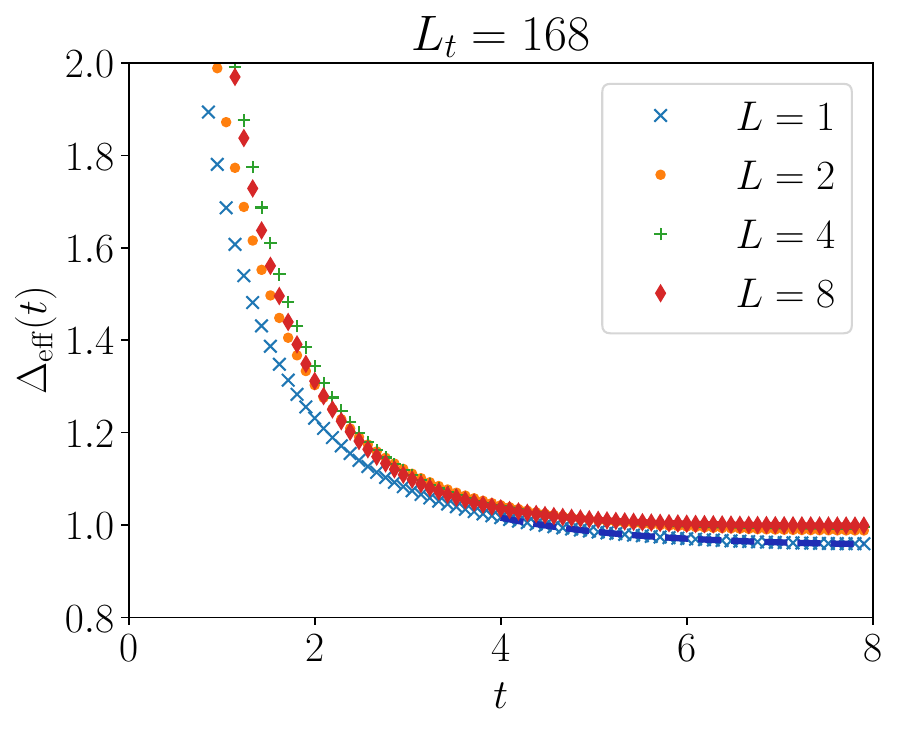}
    \caption{
    The effective dimension~\eqref{eq:def_eff_mass} for $L_t=168$ and $L=1,2,4,8$. 
    The fit curves are indicated by dashed lines.
    }
    \label{fig:meff_fit}
\end{figure}
We set $T = 16$ and take the lattices of $L=1,2,4,8$ and $L_t = 64, 96, 120, 144, 168$ that satisfy the condition $\bar{a}_s/a_t \geq 4/3$ (see Table~\ref{tab:lattices_fermion}) to obtain $\Delta_0=\Delta_0(\bar{a}_s, a_t)$. 
\begin{table}[hbt]
    \centering
    \begin{tabular}{|c|c|c|c|c|c|}\hline
        $L\backslash L_t$ & 64 & 96 & 120 & 144 & 168\\\hline
        1 & $(\circ)$ & $(\circ)$ & $(\circ)$ & $(\circ)$ & $(\circ)$ \\\hline
        2 & $(\circ)$ & $(\circ)$ & $\circ$ & $\circ$ & $\circ$\\\hline
        4 & & $(\circ)$ & $\circ$ & $\circ$ & $\circ$\\\hline
        8 & & & & $\circ$ & $\circ$ \\\hline
    \end{tabular}
    \caption{The lattices that satisfy the condition for the doublers: $\bar{a}_s/a_t \geq 4/3$ [see Eq.~\eqref{eq:overlapM}] for $T=16$.
    The lattices with the brackets are not included in the fit because they are too coarse to show the $O(a^2)$ scaling.
    }
    \label{tab:lattices_fermion}
\end{table}
The fit range is chosen to be $4\leq t<8=T/2$.
The obtained $\Delta_0(\bar{a}_s, a_t)$ are then extrapolated to the continuum limit with the quadratic ansatz:
\begin{align}
    \Delta_0(\bar{a}_s, a_t)
    =
    \Delta_0^{\rm cont}
    +
    c_s \bar{a}_s^2
    +
    c_t a_t^2,
\end{align}
where the linear terms are forbidden by the parity and time-reversal symmetries.
The scaling result is shown in Fig.~\ref{fig:fits_fermion}, which exhibits the quadratic scaling.
\begin{figure*}[htb]
    \centering
    \includegraphics[width=0.43\linewidth]{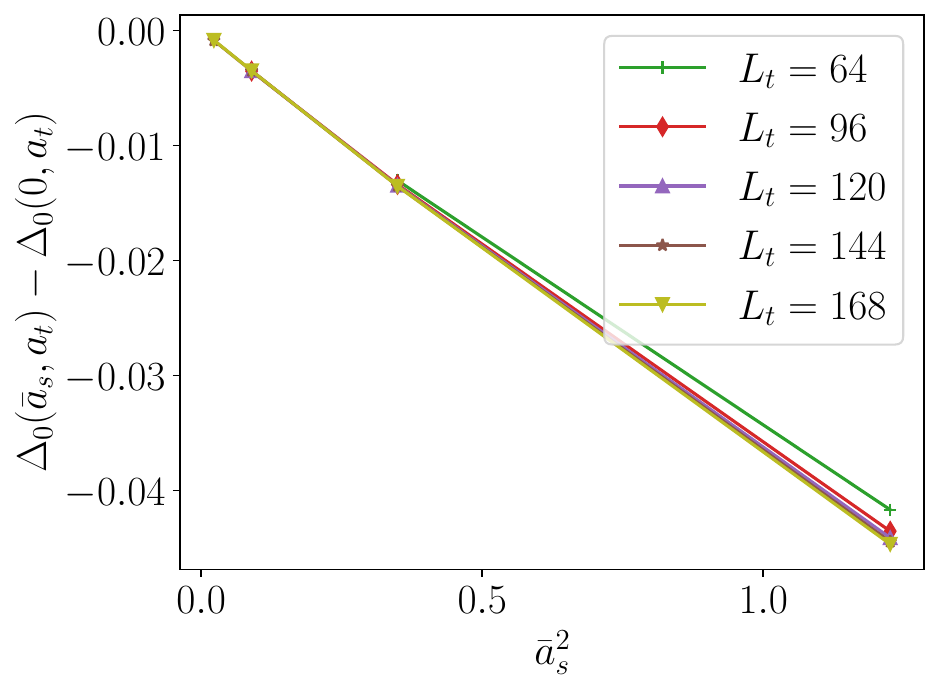}
        \hspace{0.05\linewidth}
    \includegraphics[width=0.43\linewidth]{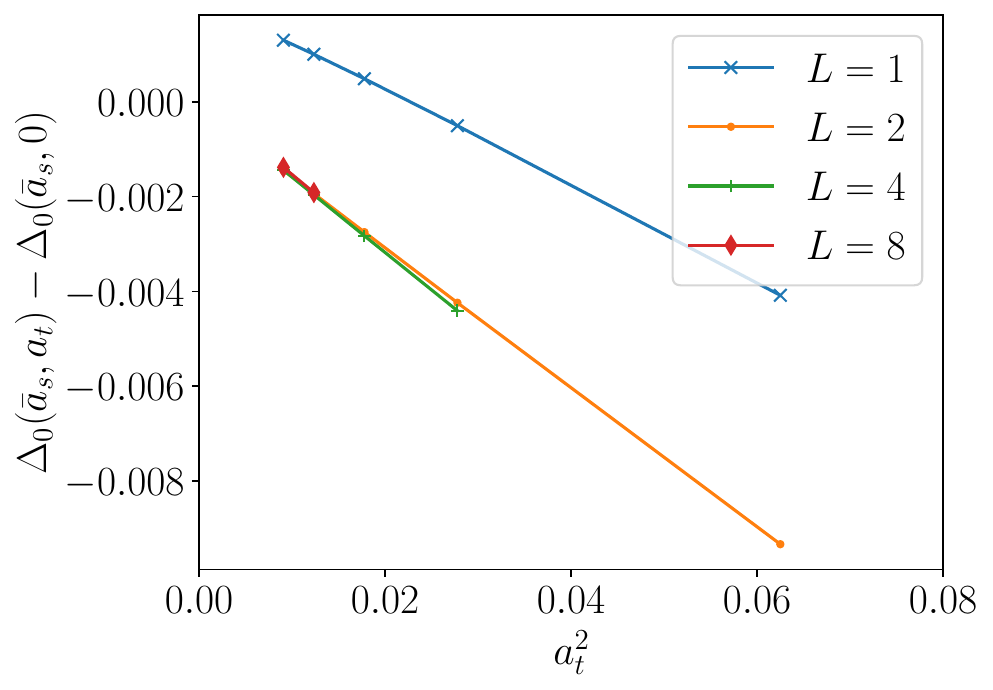}
    \caption{
    The quadratic scaling for the lowest operator dimension $\Delta_0(\bar{a}_s, a_t)$ in the overlap propagator with respect to (Left) the spatial lattice spacing and (Right) the temporal lattice spacing.
    The coarse lattices with $L_t=64,96$ or $L=1$ are not included in the final extrapolation.
    We see that the scaling curves as we decrease the lattice spacings.
    }
    \label{fig:fits_fermion}
\end{figure*}
The final estimate for $\Delta_0^{\rm cont}$ is
\begin{align}
    \Delta_0^{\rm cont}
    \approx
    0.999998(34)_{\rm sys},
\end{align}
which is in good agreement with the analytic value $\Delta_0^{\rm cont}=1$.
The coarse lattices with $L_t=64,96$ or $L=1$ are not included to obtain the central value, while the systematic error is estimated by including $L_t=96$ in the final fit.

The obtained lowest dimension $\Delta_0$ for the coarsest spatial lattice $L=1$ with $L_t = 168$ is $\Delta_0=0.953918$, whose deviation from the continuum value is less than 5\%.
The small deviation in the exponent allows us to estimate the number of excited states (or descendants) that are reproduced for a given refinement $L$ heuristically by comparing the numerical correlators with the analytic correlator $G(t; n_{\rm max})$ with a cutoff in the summation: $0\leq n\leq n_{\rm max}$.
By allowing the normalization $C$ to vary:
\begin{align}
     G^{(1,1)}_{\rm{ov}}(t)
     \simeq 
     C \cdot G^{(1,1)}(t; n_{\rm max}),
\end{align}
we perform a least-squares fit to determine the cutoff $n_{\rm max}(L)$ that minimizes the residual. 
For $T=12$ and $L_t=168$, we obtain $n_{\rm max}(1)=6$, $n_{\rm max}(2)=10$, $n_{\rm max}(4)=19$, and $n_{\rm max}(8)=32$.
The residual per degrees of freedom (DOF) is 0.028, 0.012, 0.0039, 0.038, respectively, where DOF=168-2. 
The small residual shows that, although the states around $n_{\rm max}$ are at the lattice cutoff scale and thus can be fuzzy, the simple truncation ansatz describes the lattice system reasonably.

Finally, we study the spacetime symmetries of the lattice propagators.
In Fig.~\ref{fig:prop_asym}, we compare the temporal lattice propagators $G_{\rm lat}(t)$ with $L=2$, $T=12$, $L_t=168$.
Again, the $(1,1)$ spinor component is shown.
\begin{figure}[htb]
    \centering
    \includegraphics[width=0.85\linewidth]{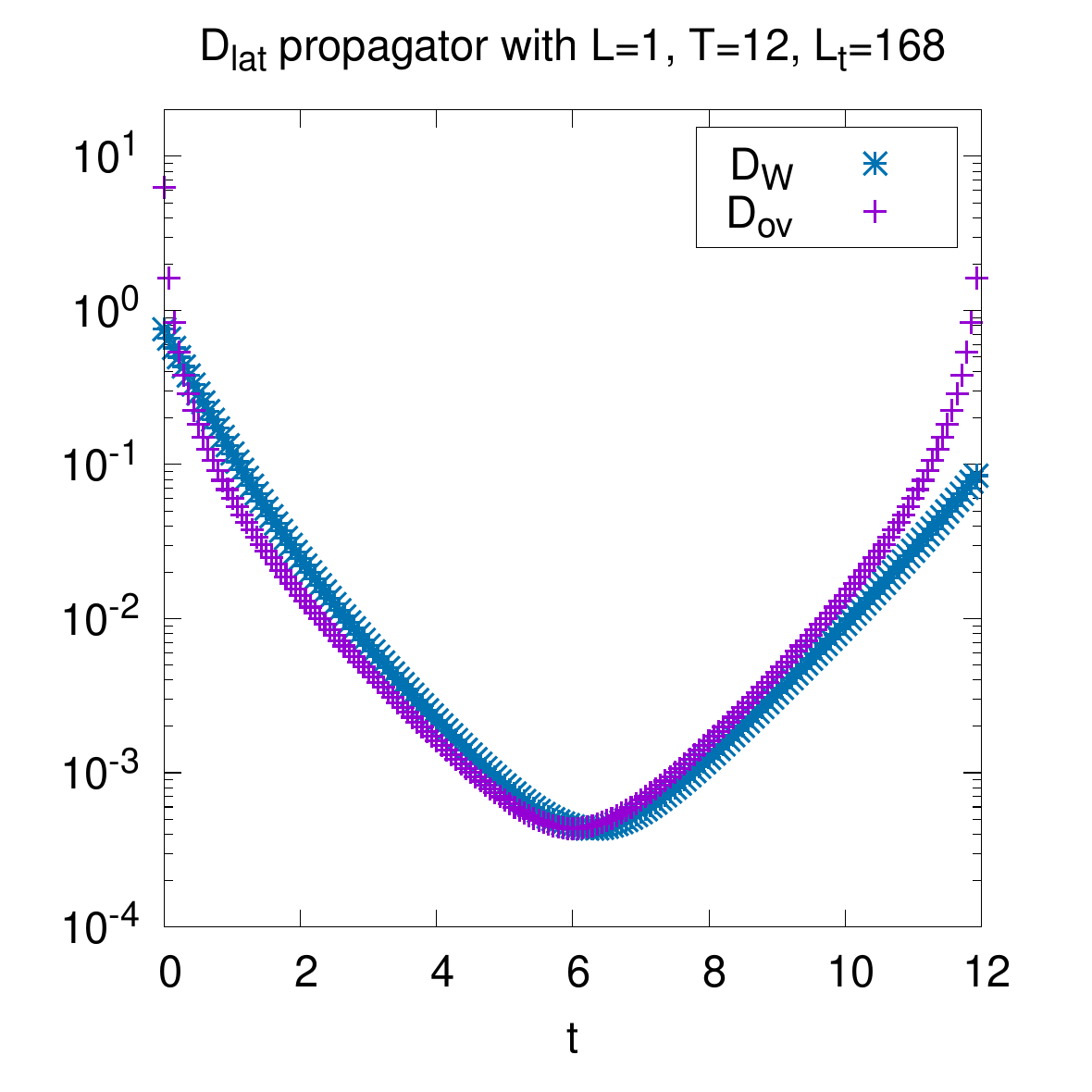}
    \caption{Comparison of the temporal propagator between the Wilson and overlap fermions with $L=1$, $T=12$, $L_t=168$.
    The breaking of the time reversal symmetry for the propagator is observed for the Wilson fermion. 
    For the overlap fermion, the Ginsparg-Wilson relation protects the symmetry in the bulk as shown in Eq.~\eqref{eq:parity_overlap_2cmp} for the $P$ symmetry.
    }
    \label{fig:prop_asym}
\end{figure}
The overlap fermion satisfies the time-reversal symmetry $T$ in the bulk thanks to the Ginsparg-Wilson relation [see Eq.~\eqref{eq:parity_overlap_2cmp} for the $P$ counterpart].
On the other hand, we observe a visible violation of the time-reversal symmetry for the Wilson fermion due to a finite lattice cutoff.
Since the global symmetries provide important quantum numbers of the theory, the result suggests that the use of the overlap fermion is strongly preferred.

\subsection{Current correlator for the gauge sector}
\label{sec:JJcorr}

For the gauge sector, we calculate the correlator of the conserved current:
\begin{align}
    J^\rho\equiv
    \frac{1}{2}
    \epsilon^{\rho\mu\nu}
    F_{\mu\nu}.
    \label{eq:Jrho}
\end{align}
In particular,
\begin{align}
    J^t
    &=
    \frac{1}{\sqrt{g}}
    F_{\theta \phi}.
    \label{eq:Jt}
\end{align}
From Eq.~\eqref{eq:approx}, we can identify the lattice operator for $J^t$ to be
\begin{align}
    J^t 
    \simeq
    \frac{1}{A_\triangle}
    \sum_{m \in p(\triangle, t)}
    \eta_m \theta_m
    \equiv J^t_{\rm lat}.
\end{align}
The analytic continuum expression for the temporal two-point function of $J^t$ is derived in Appendix~\ref{sec:gauge_prop} to be
\begin{align}
    G_g(t) &\equiv 
    \frac{1}{g^2}
    \langle
    J^t(0,0,t)
    J^t(0,0,0)
    \rangle 
    \label{eq:JJ_t_def}\\
    &=
    \frac{1}{8\pi}
    \sum_{n\geq 1}
    \sqrt{n(n+1)}(2n+1)e^{-\sqrt{n(n+1)}\vert t \vert}.
    \label{eq:exact_JJ}
\end{align}
The non-integer spacing in the exponent for the conserved current implies that the pure gauge theory ($N_f=0$) is not conformal.

The corresponding lattice correlator:
\begin{align}
    G_{g, \rm{lat}}(t) \equiv \frac{1}{g^2}\langle J^t_{\rm lat}(0,0,t) J^t_{\rm lat}(0,0,0) \rangle
    \label{eq:lattice_gauge_corr}
\end{align}
can be evaluated without Monte Carlo simulation because the gauge action~\eqref{eq:lattice_gauge} is of the Gaussian form: $S_{g, {\rm lat}} = (1/2) \theta_m M_{mn} \theta_n$.
Note, however, that the zero modes associated with the gauge symmetry exist, which parametrize the gauge-orbit direction of the configuration space.
For gauge-invariant expectation values, the zero modes are canceled between the numerator and the denominator because both the action and the operator do not have an overlap with the zero modes.
Therefore, knowing that the correlator~\eqref{eq:lattice_gauge_corr} is gauge invariant, for each pair of $(\theta_m, \theta_n)$ in the correlator, we can assign the element $\tilde{M}^{-1}_{mn}$ of the pseudo-inverse matrix $\tilde M^{-1}$, which is the inverse of $M$ in the subspace without the zero modes. 
This subspace can be understood as the quotient space where the original lattice configuration space is divided by the gauge symmetry.
This assignment does not contradict Elitzur's theorem \cite{Elitzur:1975im} because the cancellation between the numerator and the denominator does not hold for a non-gauge-invariant operator in the first place.

The projection can be performed numerically in combination with the standard conjugate gradient (CG) algorithm. By setting the initial guess for the solution $\chi$ to be $\chi_0 \equiv 0$, CG derives a polynomial $p(M)$ that represents the inverse matrix for a given source vector $b$ in the form:
\begin{align}
    \chi
    =
    \tilde{M}^{-1} P \, b
    \approx 
    p(M)\, b
    \equiv
    (p_1 M + p_2 M^2 + \cdots) \, b,
    \label{eq:main_inversion}
\end{align}
where we formally insert the projector $P$ that eliminates the zero modes of $M$.
(Since the multiplication of $M$ removes the zero modes by definition, the Krylov space does not include the zero modes.)
For the residual to converge to zero, however, the zero modes need to be projected out from the source vector in advance.
This can be numerically performed again with CG by a prior multiplication of $M$:
\begin{align}
    P \, b = \tilde{M}^{-1} M \, b.
    \label{eq:source_condition}
\end{align}
Therefore, as advertised, we can assign a numerical value to $\langle \theta_m \theta_n \rangle$ in the projected quotient space by locating the point source $b_{(n)}$ on the link $n$, whose $n$-th component is one and otherwise zero, perform the projection~\eqref{eq:source_condition} by using CG for the inversion: $b'_{(n)} \equiv P \, b_{(n)}$, and perform the main inversion again with CG: $\chi = \tilde{M}^{-1} \, b'_{(n)}$. 
The $m$-th component of the solution vector, $\chi_m$, gives the desired two-point function.

Figure~\ref{fig:jtjt} shows the thus-obtained correlator $G_{g, \rm{lat}}(t)$ in comparison to the exact formula~\eqref{eq:exact_JJ}.
\begin{figure}[htb]
    \centering
    \includegraphics[width=0.85\linewidth]{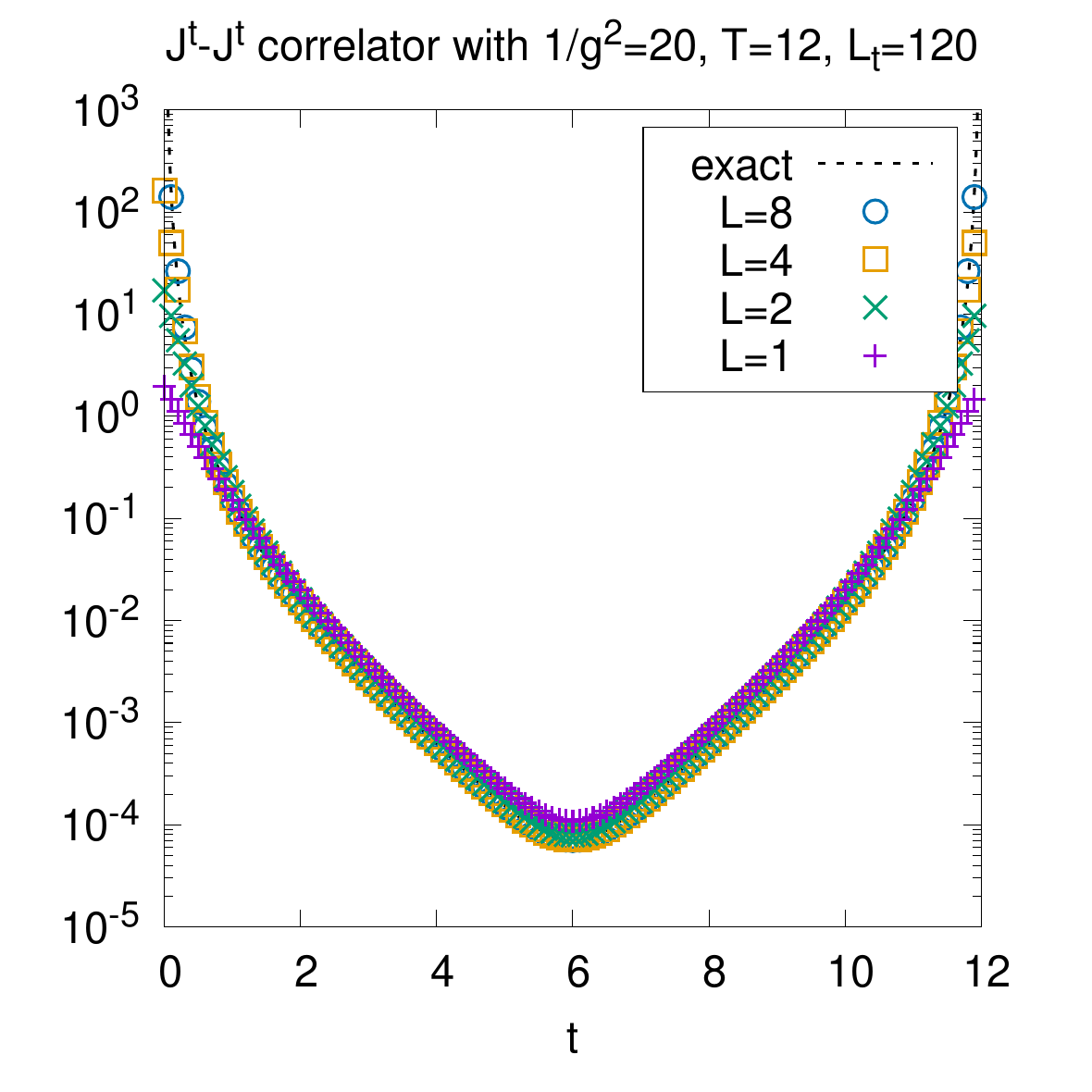}
    \caption{The gauge current correlator $G_{g}(t) = (1/g^2)\langle J^t(0,0,t)J^t(0) \rangle$ for $1/g^2=20$, $T=12$, and $L_t=120$.
    The refinement level is varied as $L=1,2,4,8$.
    }
    \label{fig:jtjt}
\end{figure}
We see a good agreement including the overall normalization, where the higher-dimensional states in the short time range are supplied as we increase the refinement level $L$. 

We perform the same procedure as in Sec.~\ref{sec:dirac_propagators} to estimate the lowest operator dimension $\Delta_0$ in the correlator with the lattices of $L=1,2,4,8$ and $L_t=24, 48, 64, 96, 120$ with $T=16$.
The quadratic scaling is again confirmed in Fig.~\ref{fig:fits_gauge}.
\begin{figure*}[hbt]
    \centering
    \includegraphics[width=0.43\linewidth]{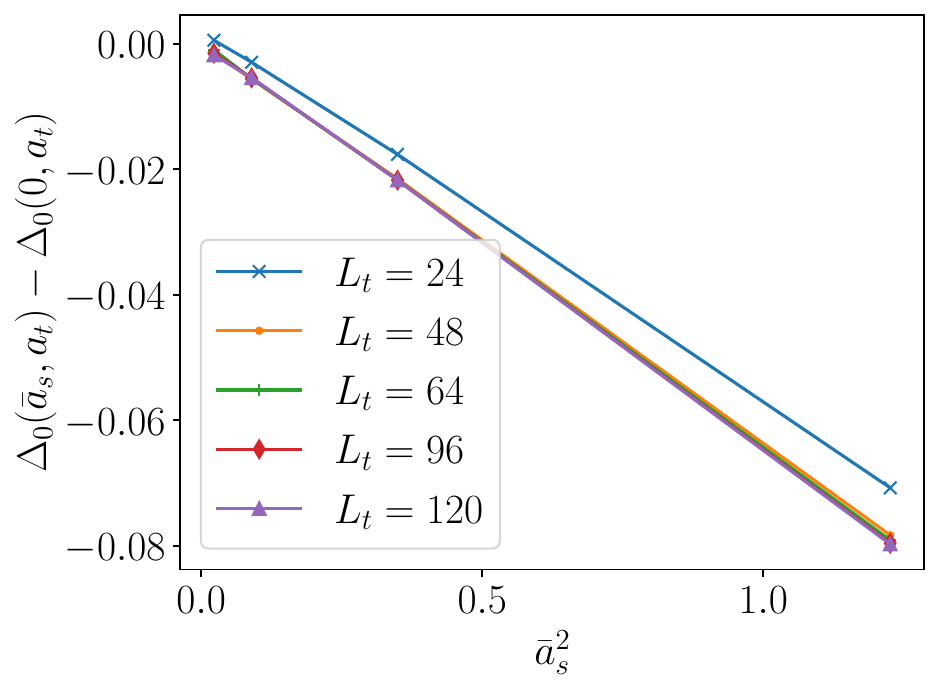}
    \hspace{0.05\linewidth}
    \includegraphics[width=0.43\linewidth]{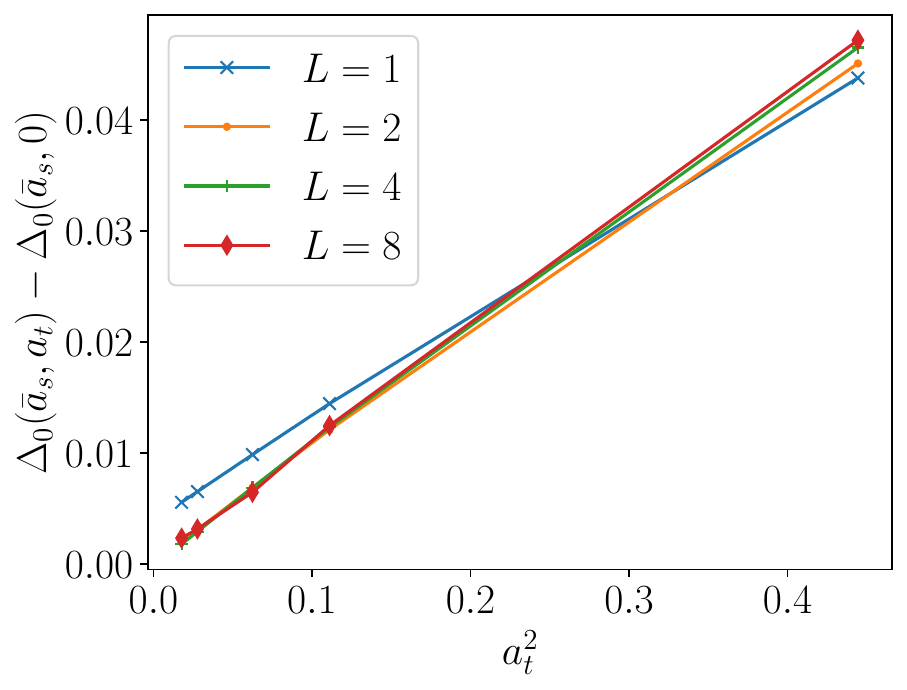}
    \caption{
    The quadratic scaling of the lowest operator dimension $\Delta_0(\bar{a}_s, a_t)$ in the gauge current correlator $G_{g,{\rm lat}}(t)$ with respect to (Left) the spatial lattice spacing and (Right) the temporal lattice spacing.
    We again observe the approach to the universal lines as we decrease the lattice spacings.
    }
    \label{fig:fits_gauge}
\end{figure*}
The final estimate for the continuum limit $\Delta_0^{\rm cont}$ is
\begin{align}
    \Delta_0^{\rm cont}
    =
    1.41409(18)_{\rm sys},
\end{align}
which is in good agreement with the analytic value $\sqrt{2}=1.41421\cdots$.
The coarse lattices with $L_t=24$ or $L=1$ are not included to obtain the central value, while the systematic error is estimated by including $L=1$ in the final fit. 

The obtained $\Delta_0(\bar{a}_s, a_t)$ for $L=1$ and $L_t=120$ is $1.33242$, whose deviation from the continuum value is less than 6\%.
We thus again heuristically estimate the number of reproduced excited states by varying the maximum integer $n_{\rm max}(L)$ in the summation in the analytic formula~\eqref{eq:exact_JJ}. 
For $T=12$ and $L_t=120$, the result is $n_{\rm max}(1)=3$, $n_{\rm max}(2)=8$, $n_{\rm max}(4)=18$, and $n_{\rm max}(8)=35$. 
The residual per DOF is
0.0031, 0.0023, 0.0031, 0.0037, respectively, where DOF=120-2.
Together with the fermion result, we thus expect a qualitative CFT study to be feasible for low-dimensional operators with the coarse lattices.

\section{Discussion}
\label{sec:discussion}

In this paper, we study the free limit of QED$_3$ in radial quantization on the lattice. 
The lattice action is constructed and confirmed to give the correct continuum limit by comparing the lattice correlators to the derived analytic formulas.
The $O(a^2)$-scaling for the lowest operator dimension in the correlators supports the claim.
By using the overlap fermion, we further show that the important global symmetries can be preserved on the lattice in the radial picture.
Having established a theoretical ground in the free limit, we can readily start calculations for the interacting theory. 
Hybrid Monte Carlo calculations are running for $N_f=2,4,6$ with the overlap fermion, and a crude conformal window study will follow.
The development of high-performance code is further in progress with Grid \cite{Boyle:2015tjk}.

It is known that anisotropic lattices require a non-perturbative fine-tuning of the coupling constants \cite{Karsch:1982ve,Burgers:1987mb} (see also Refs.~\cite{Edwards:2008ja,HadronSpectrum:2008xlg}).
In our case, a condition for the fine-tuning will be to require the full spherical symmetry (see Refs.~\cite{Brower:2020jqj,Brower:2024otr} for improvement schemes).
We consider machine-learning-based optimizations \cite{BFSM2025} to be helpful at large refinement levels $L$ to deduce a large number of coupling constants from the correlator data.

Once the coupling constants are given in a systematically improvable way, a quantitative first-principles study becomes possible.
We expect from the free limit result that the first few descendant states can be reproduced on the coarse lattices.
In fact, in the fermion propagator, the states with $0 \leq n\leq 6$ are reproduced without refinement $(L=1)$ and those with $0 \leq n\leq 10$ with the first refinement $(L=2)$, while in the gauge current correlator,  $1 \leq n\leq 3$ for $L=1$ and $1 \leq n\leq 8$ for $L=2$.
In this regard, it will be interesting to compare lattice results with the fuzzy sphere formulation \cite{Zhu:2022gjc,Fardelli:2024qla,Lauchli:2025fii}, where the number of states in the system is controlled differently and the descendants can be filled exactly up to a given level. 
The complementary advantages of the two methods can help probe unexplored conformal theories in higher dimensions.

Another interesting direction is to consider the odd-flavor theory. 
As mentioned in Introduction, the theory is considered to induce a Chern-Simons term in the effective action that violates parity \cite{Redlich:1983kn,Redlich:1983dv,Narayanan:1997by,Kikukawa:1997qh}. 
Despite the fact that the odd number of flavors results in a sign-indefinite determinant on the lattice, since the parity anomaly can be seen as a counterpart of the chiral anomaly in four dimensions, clarifying the theoretical picture as well as its phenomenological consequences would be helpful to understand quantum field theory from non-perturbative perspectives.

To study the interacting theory (regardless of even or odd $N_f$) in full theoretical rigor, it is attractive to use the domain-wall formulation \cite{Kaplan:1992bt,Shamir:1993zy,Furman:1994ky,Neuberger:1997bg} (see Refs.~\cite{Hands:2015dyp,Hands:2015qha} for the study in 3D).
Though the block-diagonal four-component overlap operator ${\cal D}_{\rm ov}$ can be obtained as a limit of the domain-wall fermion with the Wilson kernel ${\cal D}_W$ in M\"obius formulation \cite{Brower:2012vk}, it is not straightforward to single out the two-component overlap operator $D_{\rm ov}$ for the interacting theory under finite $L_s$ (the extent of the fictitious warped direction) due to an off-diagonal gamma matrix $\Gamma = \gamma_4$ or $\gamma_5$ in the hopping term in the warped direction (unless with a trick such as taking the square root).
Recently, it was proposed to formulate a chiral fermion by utilizing nontrivial manifolds \cite{Clancy:2024bjb,Kaplan:2024ezz} (see also Ref.~\cite{Aoki:2022aez,Golterman:2024ccm,Yamamoto:2025ehe}).
It is therefore interesting to further introduce nontrivial geometry in the warped direction to define the system in the domain-wall formulation.
For exploratory studies, on the other hand, the quenched calculation with twisted-mass Wilson fermion \cite{Frezzotti:2000nk,Frezzotti:2003ni} can be a viable option to remedy the breaking of chiral and accompanying discrete spacetime symmetries.

Studies along the above lines are in progress and will be reported elsewhere.

\section*{Acknowledgments}

The authors thank Cameron V.~Cogburn, A.~Liam Fitzpatrick, Evan K.~Owen, Curtis T.~Peterson, Robert D.~Pisarski, and Yuan Xin for valuable discussions. 
This work is supported in part by the Scientific Discovery through Advanced Computing (SciDAC) program under FOA LAB-2580
funded by U.S. Department of Energy, Office of Science, U.S. DOE grant No.~DE-SC0015845, and U.S. DOE grant No.~DE-SC0019139.
This document was prepared by the Quantum Finite Element (QFE) project using the resources of the Fermi National Accelerator Laboratory (Fermilab), a U.S. Department of Energy, Office of Science, Office of High Energy Physics HEP User Facility. Fermilab is managed by FermiForward Discovery Group, LLC, acting under Contract No.~89243024CSC000002.
The computation was partly performed on the Shared Computing Cluster (SCC), which is administered by Boston University's Research Computing Services. 
The authors acknowledge the Research Computing Services group for providing support.
The code and data that support the findings of this paper are publicly available \cite{nobuyuki_matsumoto_2025_17210221}, which uses the newQFE package \cite{newQFE} to generate the refined icosahedron.

\appendix

\section{Conventions}
\label{sec:fermion}

On a generic manifold in $D$ dimensions, the vierbein $e_\mu^a$ can be taken such that the local Lorentz space has the flat metric:
\begin{align}
  e_{\mu}^a \, e_{\nu}^b \, g^{\mu \nu} = \delta^{ab}.
\end{align}
We define the invariant volume form as
\begin{align}
    dV 
    &\equiv 
(1/D!)
\,
\epsilon_{\mu_1\cdots\mu_D}\,
dx^{\mu_1}\wedge 
\cdots \wedge
dx^{\mu_D}
\\    
&=
e^1_{\mu_1}
    \cdots
    e^D_{\mu_D}
    dx^{\mu_1}\wedge 
\cdots \wedge
dx^{\mu_D},
\end{align}
where the antisymmetric tensor $\epsilon_{\mu_1\cdots\mu_D}$ includes the volume factor $\sqrt{g}$. 
For example, for $S^2\times \mathbb{R}$,
\begin{align}
    \epsilon_{\theta \phi t}
    =
    \det e^{a}_\mu
    =
    \sqrt{g}.
\end{align}

We write as $M^R_{ab}$ the generators of the local Lorentz group $SO(D)$ in the representation $R$, which we take to be antihermitian.
The covariant derivative acting on the field $\phi^R$ in the representation $R$ is then
\begin{align}
    &\nabla_\mu^R \phi^R \equiv (\partial_\mu 
    + 
    \omega^R_\mu) \phi^R,
    \label{eq:spinor_cov}
    \\
    &{\omega}^R_\mu 
    \equiv
    \frac{1}{2}
    \omega_\mu^{ab} M^R_{ab}.
\end{align}
We define the (plain) covariant derivative $\nabla_\mu$ to act on both vector and local Lorentz indices.
The vierbein satisfies the parallel transport equation:
\begin{align}
  \nabla_\mu e^a_\nu
  =
  \partial_\mu e^a_{\nu}
  -
  \Gamma_{\mu\nu}^\gamma e_\gamma^a
  +
  \omega_{\mu\, b}^a e_\nu^b
  = 0.
  \label{eq:parallel_tetrad}
\end{align}

Along a geodesic $\gamma_{xy}$ with the proper length $s$, the parallel transport equation
\begin{align}
    &e_{xy}^\mu \nabla_\mu^R \phi^R 
    = 0,
\end{align}
where $e_{xy}^\mu \equiv dx^\mu/ds$, can be integrated to give
\begin{align}
    \phi^R(x) 
    = \Omega_{xy}^R \phi^R(y).
    \label{eq:parallel_transport_R}
\end{align}
$\Omega_{xy}^R$ is the Wilson line:
\begin{align}
  \Omega_{xy}^R
  \equiv
  {\cal P} \exp
  \left(
  \int_{\gamma_{xy}} ds \,
  e_{xy}^\mu
  \omega_\mu^R
  \right),
  \label{eq:int_vec_ll}
\end{align}
where ${\cal P}$ stands for the path-ordering that runs from $x$ to $y$ along $\gamma_{xy}$ from left to right, which is the direction of increasing $s$.

For a spinor representation $R=S$, we introduce the gamma matrices $\gamma_a$ that satisfy $\{ \gamma_a, \gamma_b \} = 2\delta_{ab}$. 
In the main text, we have two representations in this class, $R=S, S'$, where we take $\sigma_a$ as the generators for $R=S$ and the block-diagonal $\gamma_a$ given in Eq.~\eqref{eq:def_gamma} as the generators for $R=S'$.
The corresponding $SO(D)$ generators are
\begin{align}
    M_{ab}^S \equiv \frac{1}{4}[\gamma_a,\gamma_b].
\end{align}
From $e_{xy}^\mu \nabla_\mu e_{xy}^\nu = 0$ and $\nabla_\mu \tensor{(\gamma_a)}{^i_j} = 0$, one can show that
\begin{align}
    \gamma_a 
    e_{xy}^a(x) 
    \Omega^S_{xy}
    =
    \Omega^S_{xz}
    \gamma_a 
    e_{xy}^a(z)
    \Omega^S_{zy}
    \quad
    (z \in \gamma_{xy}).
    \label{eq:tetrad_hypo1}
\end{align}
In particular,
\begin{align}
    \gamma_a 
    e_{xy}^a(x) 
    \Omega^S_{xy}
    =
    \Omega^S_{xy}
    \gamma_a 
    e_{xy}^a(y),
    \label{eq:tetrad_hypo2}
\end{align}
referred to as the tetrad hypothesis in Ref.~\cite{Brower:2016vsl}.
$\Omega_{xy}^S$ will be written simply as $\Omega_{xy}$.

\section{Calculation of the lattice spin connection \texorpdfstring{$\Omega_{y_1 y_2}$}{}}
\label{sec:geometry_s2}

In this appendix, we summarize a method to analytically calculate the lattice spin connection matrices $\Omega_{y_1 y_2}$ on $S^2$.
Starting from the geodesic equations:
\begin{align}
  &\tan(\phi-\phi_0) = \iota_\phi k \tan(s-s_0), 
  \label{eq:solution_geodesic1}\\
  &\cos \theta = -\iota_\theta \sqrt{1-k^2} \sin (s-s_0),
  \label{eq:solution_geodesic2}
\end{align}
one can eliminate $s-s_0$ to obtain
\begin{align}
  \frac{1}{\tan^2 \theta}
  =
  \Big( \frac{1}{ k^2} -1 \Big)
  \sin^2 (\phi-\phi_0).
  \label{eq:geodesic_nos}
\end{align}
The constants $k>0$ and $\phi_0\in[0,2\pi)$ parameterize the shape of the geodesic. 
The sign factor $\iota_\phi\equiv\pm1$ determines the sign of $\dot \phi$, and the sign factor $\iota_\theta\equiv \pm1$ in Eq.~\eqref{eq:solution_geodesic2} determines the sign of $\dot \theta$.
Given the shape parameters and the directional sign factors, $s_0$ determines the initial point of the curve, while the endpoint of the curve is at $s = s_0 + \ell$ with the arclength $\ell$.
When $\theta$ is not monotonic, the curve needs to be divided into two patches at the extremum of $\theta$ at $s-s_0=\pm\pi/2$ mod $2\pi$.

With the analytic formula for the geodesic $\gamma_{y_1y_2}$, we can readily calculate the Wilson line~\eqref{eq:int_vec_ll}. 
In $D=2$, the path-ordering is trivial,
giving Eqs.~\eqref{eq:expn_omega} and \eqref{eq:expn_int_part}.
In practice, it is convenient to avoid coordinate singularities for numerical calculations.
For this, we tilt the polar axis slightly such that the pole points do not lie on top of the sites or the links.
Performing such a rotation corresponds to choosing a different gauge for the local Lorentz frame.
For the calculation in the main text, the integral~\eqref{eq:expn_int_part} is numerically evaluated with the GSL libary.

We remark that the coordinate singularities of $\omega^S_\alpha$ are related to the double-sheet structure of the fermion field on $S^2$.
Using the $(\theta,\phi)$ coordinates, we have for a closed curve $\gamma$ on $S^2$:
\begin{align}
    \oint_\gamma 
    dy^\alpha
    \omega^S_\alpha
    &=
    \frac{i\sigma_3}{2} 
    \Big(
    \int_{{\rm int}(\gamma)} dV
    -
    2\pi \eta(\gamma)
    \Big).
    \label{eq:expn_s2}
\end{align}
The first term in the bracket is the area of the interior ${\rm int}(\gamma)$ of $\gamma$.
The second term is from the nonzero divergence at each pole, where $\eta(\gamma)=1$ if the curve $\gamma$ encircles the north (south) pole in the positive (negative) $\phi$ direction and is 0 if $\gamma$ does not include a pole in the interior.
The additional divergence term in the local curvature formula~\eqref{eq:expn_s2} can be canceled by introducing a cut $\gamma_c$ that runs from the north pole to the south pole and does not lie on top of the lattice sites.
The antiperiodic boundary condition for $\psi$ and $\bar\psi$ is then imposed in the $\phi$-direction by assigning the minus sign when the link crosses $\gamma_c$.
The parallel-transported variable~\eqref{eq:parallel_transport_R} acquires a minus sign when it crosses $\gamma_c$, precisely canceling the divergence term.

The antiperiodic boundary condition can be reinterpreted as a sign factor for $\Omega_{y_1y_2}$. In this case, Eq.~\eqref{eq:tetrad_hypo2} and Eq.~\eqref{eq:expn_s2} without the $\eta$ term can be identified as the defining properties of the lattice spin connection. 
See Ref.~\cite{Brower:2016vsl} for more on this view and an algorithm to compute $\Omega_{y_1y_2}$ as an optimization problem.

\section{Derivation of the continuum correlators}
\label{sec:derivation}

In this appendix, we derive the analytic expressions for the correlators quoted in Sec.~\ref{sec:tests}.
The free propagators for the fermion and gauge fields can be expressed by the eigenvalues and the eigenfunctions of the kernel matrices.
The line of argument for solving the eigenvalue problem follows Ref.~\cite{Abrikosov:2002jr}, which addressed the Dirac equation on $S^2$. 
For completeness, we provide a full explanation in our context with corrections to the fermion result on $S^2$.
Note that the two Killing vectors, $\partial_\phi$ and $\partial_t$, commute with each other and
with the kernel matrices for the propagators, and thus they can be simultaneously diagonalized. 
The solutions are obtained in the simultaneously diagonalizing basis as they resolve the degenerate eigenvalues.

\subsection{Free fermion propagator}
\label{sec:ferm_prop}

To invert the free Dirac matrix $D\equiv \sigma^a e_a^\mu {\nabla}^S_\mu$, we solve the eigenvalue problem:
\begin{align}
    D \psi
    =
    i \lambda \psi.
    \label{eq:D}
\end{align}
Note that the definition of $\lambda$ differs from the main text by a factor of $i$.
The Dirac matrix is antihermitian with respect to the inner product:
\begin{align}
    (\psi_1, \psi_2)
    \equiv
    \int
    dV \,
    \psi_1^\dagger(x) 
    \psi_2(x),
\end{align}
which makes $\lambda\in\mathbb{R}$.
The eigenfunctions of $D$ are a subset of the eigenfunctions of the square of the Dirac matrix, which can be rewritten with the Lichnerowicz formula:
\begin{align}
    D^2 \psi
    =
    \Big(
    g^{\mu\nu} \nabla_\mu \nabla^S_\nu \psi
    -
    \frac{1}{4}
    R
    \Big) \psi,
    \label{eq:Dsq}
\end{align}
where $R$ is the Ricci scalar.
We first consider the eigenvalue problem for the squared matrix, which reads on $S^2 \times \mathbb{R}$:
\begin{align}
    &\Big[
    \partial_t^2
    +
    \partial_\theta^2
    +
    \frac{1}{\sin^2\theta}
    \Big(\partial_\phi 
    - \frac{i}{2}\cos\theta\sigma_3
    \Big)^2
    +
    \cot\theta \partial_\theta
    -
    \frac{1}{2}
    \Big] \psi
    \nonumber\\
    & \hspace{0.65\linewidth}
    =
    -\lambda^2 \psi.
    \label{eq:dirac_sq}
\end{align}

We first drop the $t$-dependence and derive the solution on $S^2$.
We set $z \equiv \cos\theta$ and take the ansatz:
\begin{align}
    \psi(\theta, \phi)
    =
    e^{im\phi} u_m(z),
\end{align}
where $m \in \mathbb{Z} + 1/2$ since $\psi$ is antiperiodic in $\phi$ (see Appendix~\ref{sec:geometry_s2}).
Equation~\eqref{eq:dirac_sq} reduces to
\begin{align}
    &\Big[
    \frac{d}{dz}
    (1-z^2)
    \frac{d}{dz}
    -
    \frac{m^2 - \sigma_3 m z + z^2/4}{1-z^2}
    + \lambda^2 - \frac{1}{2}
    \Big]
    u_m(z)
    \nonumber\\
    & \hspace{0.7\linewidth}
    = 0.
    \label{eq:eq_u_z}
\end{align}
The expression in the bracket is diagonal, and the differential equation can be solved independently for each component. 
Writing the eigenvalues of $\sigma_3$ as $\iota_3 \equiv \pm 1$ and separating the sign part of $m$ as $m \equiv \iota_m \vert m \vert $, we see that each component obeys the same differential equation with the variable $\iota_3 \iota_m z$ for a given $\vert m \vert$.
We write the solution as $\xi_{\vert m \vert} (\iota_3 \iota_m z)$.
By factoring out the singular part, the differential equation for $\xi_{\vert m \vert}(z)$ reduces to a hypergeometric form, and the relevant solution can be easily found:
\begin{align}
    \xi_{\vert m\vert}(z)
    &\propto
    (1-z)^{\frac{1}{2}( \vert m \vert - 1/2)}
    (1+z)^{-\frac{1}{2}( \vert m \vert + 1/2)}
    \nonumber\\
    & \hspace{0.1\linewidth}
    \times F\Big(
    -\vert \lambda \vert,
    \vert \lambda \vert;
    \frac{1}{2}+\vert m\vert;
    \frac{1 - z}{2}
    \Big),
\end{align}
where $F(a,b;c;x)$ is the hypergeometric function.
We impose regularity at $z=-1$, whose value can be calculated by Gauss's theorem:
\begin{align}
    F (a,b;c; 1 )
    =
    \frac{\Gamma(c) \Gamma(c-a-b)}
    {\Gamma(c-a)\Gamma(c-b)}.
\end{align}
The condition implies
\begin{align}
    \frac{1}{2}+ \vert m \vert 
    - \vert\lambda\vert
    \equiv - n,
    \quad
    n \in \mathbb{Z}_{\geq 0},
\end{align}
which leads to the expression
\begin{align}
    \xi_{\vert m\vert,n}(z)
    &=
    (1-z)^{\frac{1}{2}( \vert m \vert - 1/2)}
    (1+z)^{-\frac{1}{2}( \vert m \vert + 1/2)}
    \nonumber\\
    & \hspace{0.1\linewidth}
    \times P^{(\vert m\vert-1/2, -\vert m\vert-1/2 )}_{n+\vert m\vert+1/2}
    (z),
\end{align}
where $P_j^{(\alpha,\beta)}(z)$ is the Jacobi polynomial:\footnote{
    Conventionally, Jacobi polynomials are defined for $\alpha,\beta>-1$ to ensure the integrability of the weight function, which is not the case in our usage.
    Nevertheless, Rodrigues' formula is still valid in our application:
    \begin{align}
    P_j^{(\alpha,\beta)}(z)
    &=
    \frac{(-1)^j}{2^j j!}
    (1-z)^{-\alpha}
    (1+z)^{-\beta}
    \frac{d^j}{dz^j}
    \Big[
    (1-z)^{\alpha+j}
    (1+z)^{\beta+j}
    \Big]
    \label{eq:rodrigues}
\end{align}
because $j$, $\alpha+j$, $\beta+j$ are nonnegative integers in all occasions.
Accordingly, well-known formulas such as normalization and orthogonality relations mostly apply in the generalized case \cite{szego75}. 
The only exception that appears in our study is in the zero modes of the gauge field, which can be easily distinguished by a singular integral. \label{fn:integrability}
}
\begin{align}
    P_{j}^{(\alpha,\beta)}(z)
    &\equiv
    \frac{\Gamma(\alpha+1+j)}{\Gamma(\alpha+1) j!}
    \nonumber\\
    & \hspace{-0.1\linewidth}
    \times
    F\Big(-j,1+\alpha+\beta+j;\alpha+1;\frac{1-z}{2}\Big).
    \label{eq:def_P}
\end{align}
From Rodrigues' formula~\eqref{eq:rodrigues}, the expansion around $z=1$ for $\alpha\geq0$ is
\begin{align}
    P_j^{(\alpha,\beta)}(z)
    &=\frac{(\alpha+j)!}{\alpha! \, j!}
    \Big(1+O(1-z)\Big),
\end{align}
and the expansion around $z=-1$ for $\beta<0$ is
\begin{align}
    P_j^{(\alpha,\beta)}(z)
    &=(-1)^{j-\beta}
    2^\beta
    \frac{(\alpha+j)!}{(-\beta)!(\alpha+\beta+j)!}
    (1+z)^{-\beta}
    \nonumber\\
    & \hspace{0.06\linewidth}
    \times
    \Big(
    1+O(1+z)
    \Big).
\end{align}
The function $\xi_{\vert m\vert}(z)$ thus simplifies at the edge points:
\begin{align}
    \xi_{\vert m \vert, n}(1) = 
    \frac{\delta_{\vert m\vert,1/2}}
    {\sqrt{2}}
    ,
    \quad
    \xi_{\vert m \vert, n}(-1) = 0.\label{eq:xi_deltam}
\end{align}

To obtain the eigenfunctions of $D$, we pair the spinor components which we have solved independently for the squared matrix in Eq.~\eqref{eq:eq_u_z}.
Note the $\sigma_3$-hermiticity:
For $D \psi(y) = i\lambda \psi(y)$, $D \sigma_3\psi(y) = -i\lambda \sigma_3\psi(y)$.
The result is (the symbol $\iota_3=\pm1$ is reused as a label)
\begin{align}
    &D \psi_{m,n,\iota_3}
    =
    \iota_3 i \lambda_{\vert m \vert,n}
    \psi_{m,n,\iota_3}, \\
    &
    \lambda_{\vert m \vert,n}
    \equiv
    n+\vert m \vert + 1/2,
\end{align}
where
\begin{align}
    &\psi_{m,n,\iota_3}(\theta,\phi)
    = 
    \frac{e^{i m \phi}}{\sqrt{2\pi}}
    \frac{1}{\sqrt{2}}
    \left(
    \begin{array}{c}
        \xi_{\vert m \vert,n}(\iota_m z) \\
        \iota_3 \iota_m i (-1)^n
        \xi_{\vert m \vert,n}(- \iota_m z)
    \end{array}
    \right).
\end{align}
The eigenfunctions are orthogonal:
\begin{align}
    (\psi_{m,n,\iota_3}, \psi_{m',n',\iota_3'})
    =
    c_{\vert m\vert,n}^2
    \,
    \delta_{mm'}
    \delta_{nn'}
    \delta_{\iota_3 \iota_3'},
\end{align}
where
\begin{align}
    c_{\vert m\vert, n}^2
    &\equiv
    \frac{1}{2n+2\vert m\vert+1} 
    \frac{n! (n+2\vert m\vert)!}
    {(n+\vert m\vert-1/2)!
    (n+\vert m\vert+1/2)!}.
    \label{eq:normalization_psi}
\end{align}

\begin{widetext}
To add the radial direction, we note that, for the free theory, the purely imaginary eigenvalues form complex conjugate pairs, though we do not have $\sigma_3$-hermiticity in three spacetime dimensions:
For $D \psi(x) = i\lambda \psi(x)$, $D \sigma_3\psi(x_T) = -i\lambda \sigma_3\psi(x_T)$.
By following the same steps, the solution can be obtained as
\begin{align}
    &D \psi_{k,m,n,\iota_3}
    =
    \iota_3 i \lambda_{\vert k\vert,\vert m \vert,n}
    \psi_{k,m,n,\iota_3}, \\
    &
    \lambda_{\vert k\vert, \vert m \vert,n}
    \equiv
    \sqrt{\vert k\vert^2 + (n+\vert m \vert + 1/2)^2}.
\end{align}
The eigenfunctions are given by
\begin{align}
    &\psi_{k,m,n,\iota_3}(\theta,\phi,t)
    \equiv
    \frac{e^{\iota_3 i kt}}{\sqrt{2\pi}}
    \frac{e^{i m \phi}}{\sqrt{2\pi}}
    \times
    \begin{cases}
    \left(
    \begin{array}{c}
        \cos(\gamma_{\vert k\vert,\vert m\vert,n}/2)
        \,
        \xi_{\vert m \vert,n}(\iota_m z) \\
        \iota_3 \iota_m i (-1)^n
        \sin(\gamma_{\vert k\vert,\vert m\vert,n}/2)
        \,
        \xi_{\vert m \vert,n}(-\iota_mz)
    \end{array}
    \right)
    \quad
    (k>0), \\
    \left(
    \begin{array}{c}
        \sin(\gamma_{ \vert k\vert,\vert m\vert,n}/2)
        \,
        \xi_{\vert m \vert,n}(\iota_m z) \\
        \iota_3 \iota_m i (-1)^n
        \cos(\gamma_{\vert k\vert,\vert m\vert,n}/2)
        \,
        \xi_{\vert m \vert,n}(-\iota_m z)
    \end{array}
    \right)
    \quad
    (k<0),
    \end{cases}
\end{align}
\end{widetext}
where the angular variable $\gamma_{\vert k\vert,\vert m\vert,n}$ is defined by
\begin{align}
    \cos \gamma_{\vert k\vert,\vert m\vert,n}
    &\equiv
    \frac{\vert k\vert}
    {\lambda_{\vert k\vert,\vert m\vert,n}}.
    \label{eq:gamma}
\end{align}
The eigenfunctions satisfy the orthogonality relation:
\begin{align}
    (\psi_{k,m,n,\iota_3}, \psi_{k',m',n',\iota_3'})
    =
    c_{\vert m\vert,n}^2 
    \,
    \delta(k-k') 
    \,
    \delta_{mm'} \delta_{nn'}
    \delta_{\iota_3 \iota_3'}.
\end{align}
By collectively writing the eigenfunctions of $D$ as $\psi_K$:
\begin{align}
    D \psi_K &= i
    \Lambda_K \psi_K,
\end{align}
and using the differential equation for the propagator $G(x,x')$:
\begin{align}
    D(x) G(x,x')
    =
    \frac{1}{\sqrt{g(x)}}
    \delta^3(x-x'),
    \label{eq:def_G}
\end{align}
the continuum propagator can be readily given as
\begin{align}
    G(x,x')=
    \sum_K
    \frac{\psi_K(x) \psi_K^\dagger(x')}{i\Lambda_K c_K^2}.
    \label{eq:Gf}
\end{align}

We calculate as an example the propagator in the temporal direction.
Because of Eq.~\eqref{eq:xi_deltam}, the expression simplifies by choosing $x=(0,0,t)$ and $x'=(0,0,0)$.
The resulting function $G(t)$ is
\begin{align}
    G(t)
    &=
    \sigma_3
    \sum_{n \geq 0} 
    \frac{n+1}{2\pi}
    \int_{-\infty}^\infty
    \frac{dk}{2\pi i}
    \frac{k e^{ikt}}{k^2 + (n+1)^2}
    \nonumber\\
    &=
    {\rm sign}(t)
    \,
    \sigma_3
    \sum_{n \geq 0} 
    \frac{n+1}{4\pi}
    \,
    e^{-(n+1)\vert t \vert},
    \label{eq:exact_Gf2}
\end{align}
as quoted in the main text.
The tower of integer-spaced exponents shows the expected spectrum of a conformal field theory with the primary operator with dimension $\Delta=1$.

\subsection{Free gauge propagator}
\label{sec:gauge_prop}

To calculate the free gauge propagator analytically, we add the gauge-fixing term:
\begin{align}
    S_g + S_{gf}
    &\equiv
    \frac{1}{g^2}
    \int 
    dV\,
    \Big(
    \frac{1}{4}
    F^{\mu\nu} F_{\mu\nu}
    +
    \frac{1}{2}
    (\nabla_\mu A^\mu)^2
    \Big) \\
    &= 
    \frac{1}{2g^2}
    \int dV\,
    A^\mu
    (-g_{\mu\nu} \nabla^\rho \nabla_\rho 
    + R_{\mu\nu}
    ) A^\nu.
\end{align}
The eigenvalue problem of our interest is
\begin{align}
    \tensor{K}{^\mu_\nu} 
    A^\nu
    =
    \lambda A^\mu,
    \label{eq:lap_A} 
    \quad
    \tensor{K}{^\mu_\nu}
    \equiv
    - 
    \tensor{\delta}{^\mu_\nu}
    \nabla^\rho \nabla_\rho    
    + \tensor{R}{^\mu_\nu}.
\end{align}
The kernel operator $\tensor{K}{^\mu_\nu}$ is hermitian with respect to the inner product:
\begin{align}
    (A_1, A_2)
    \equiv 
    \int
    dV\,
    g_{\rho\sigma} 
    (A_1^\rho)^* A_2^\sigma,
    \label{eq:inner_vector}
\end{align}
under the condition that the boundary integrals vanish upon integrating by parts. 
The Laplacian for the vector field $A^\mu$ is given in the $(\theta,\phi,t)$-coordinates by
\begin{widetext}
\begin{align}
    \nabla^\rho\nabla_\rho 
    &=
    \partial_t^2 +
    \left(
    \begin{array}{ccc}
        \partial_\theta^2
        + 
        \frac{\partial_\phi^2}{\sin^2\theta} 
        +
        \cot \theta \partial_\theta
        -
        \cot^2 \theta
        &  
        -2 \cot\theta \partial_\phi
        &
        \\
        2 \frac{\cot\theta}{\sin^2\theta} \partial_\phi
        &
        \partial_\theta^2
        + 
        \frac{\partial_\phi^2}{\sin^2\theta} 
        +
        3 \cot \theta \partial_\theta
        -
        1
        &
        \\
        &&
        \partial_\theta^2
        + 
        \frac{\partial_\phi^2}{\sin^2\theta} 
        +
        \cot \theta \partial_\theta
    \end{array}
    \right) \nonumber\\
    &=
    \partial_t^2 +
    \left(
    \begin{array}{ccc}
        \partial_z (1-z^2)\partial_z
        + 
        \frac{\partial_\phi^2}{1-z^2} 
        - 
        \frac{z^2}{1-z^2}
        &  
        -2 \frac{z}{\sqrt{1-z^2}} \partial_\phi
        &
        \\
        2 \frac{z}{(1-z^2)^{3/2}} \partial_\phi
        &
        \partial_z (1-z^2)\partial_z
        +
        \frac{\partial_\phi^2}{1-z^2} 
        -2z\partial_z
        - 
        1
        &
        \\
        &&
        \partial_z
        (1-z^2)
        \partial_z
        +
        \frac{\partial_\phi^2}{1-z^2}
    \end{array}
    \right),
\end{align}
\end{widetext}
and the Ricci tensor by
\begin{align}
    \tensor{R}{^\mu_\nu}
    =
    \left(
    \begin{array}{ccc}
        1 && \\
        & 1 &\\
        &&0
    \end{array}
    \right).
\end{align}
As in the fermion case, the curvature term provides a mass to the transverse modes, which has appeared explicitly as a result of gauge fixing. 

Again the eigenvalue problem can be solved first on $S^2$, where the eigenfunctions will be labeled by $(m, n)$ ($m\in\mathbb{Z},$ $n \in \mathbb{Z}_{\geq0}$).
The zero modes that correspond to $(m,n)=(0,0)$ will not be included in the spectrum because the boundary integral does not vanish for the operator $\tensor{K}{^\mu_\nu}$ to be hermitian (see also Footnote~\ref{fn:integrability}).
The radial direction adds the third quantum number $k \in \mathbb{R}$, where the $t$-dependence is simply the plane wave. 
The result is
\begin{align}
    &
    \tensor{K}{^\mu_\nu}
    A^\nu_{k,m,n,r}
    =
    \lambda_{\vert k\vert,\vert m\vert,n }
    A^\mu_{k,m,n,r}, 
    \\
    &\lambda_{\vert k\vert,\vert m\vert, n}
    \equiv
    k^2
    +
    (n+\vert m\vert)(n+\vert m\vert+1),
\end{align}
where $r\in\{+,-,t\}$ labels the polarizations:
\begin{align}
    \epsilon_\pm^\mu(z)
    &\equiv
    \frac{1}{\sqrt{2}}
    \left(
    \begin{array}{c}
        1\\
        \pm
        \frac{ i}{\sqrt{1-z^2}}
        \\
        0
    \end{array}
    \right),
    \quad
    \epsilon_t^\mu
    \equiv
    \left(
    \begin{array}{c}
        0
        \\
        0
        \\
        1
    \end{array}
    \right).
\end{align}
The transverse solutions are
\begin{align}
    &A^\mu_{k,m,n,\pm}
    =
    \frac{e^{i kt}}{\sqrt{2\pi}}
    \frac{e^{im\phi}}{\sqrt{2\pi}}
    Q_{\pm m}(z)
    f_{\vert m \vert, n}( \pm \iota_m z)
    \,
    \epsilon_\pm^\mu(z)
    \nonumber\\
    & \hspace{0.55\linewidth} (m\neq0, n \geq 0),
    \label{eq:sol_A_sph}
    \\
    &A^\mu_{k,0,n,\pm}
    =
    \frac{e^{i kt}}{\sqrt{2\pi}}
    \frac{1}{\sqrt{2\pi}}
    Q_{0}(z)
    f_{0, n}(z)
    \,
    \epsilon_\pm^\mu(z)
    \nonumber\\
    & \hspace{0.55\linewidth}
    (m=0,
    n\geq 1),
    \label{eq:sol_A_sph0}
\end{align}
where $\iota_m = {\rm sign}(m)$ and the functions $Q_m(z)$, $f_{\vert m \vert,n}$ are given by
\begin{align}
    Q_m(z)
    &\equiv
    (1-z)^{\frac{1}{2}(m-1)}
    (1+z)^{-\frac{1}{2}(m+1)},
    \\
    f_{\vert m \vert,n}(z) 
    &\equiv
    P_{n+\vert m\vert+1}^{(\vert m\vert-1,-\vert m\vert-1)}(z)
    \quad
    (m\neq 0 , n \geq 0),\\
    f_{0,n}(z) 
    &\equiv
    (1-z) 
    P_{n}^{(1,-1)}(z)
    \quad
    (m=0,n\geq 1).
\end{align}
As for the longitudinal polarization, we have
\begin{align}
    &A^\mu_{k,m,n,t}
    =
    \frac{e^{i kt}}{\sqrt{2\pi}}
    \frac{e^{im\phi}}{\sqrt{2\pi}}
    P_{n+\vert m \vert}^{\vert m\vert}(z)
    \,
    \epsilon_t^\mu
    \quad
     (m \in \mathbb{Z}, n \geq 0)
     ,
\end{align}
where $P_{j}^{\vert m\vert}(z)$ is the associated Legendre polynomial.
The eigenfunctions are orthogonal:
\begin{align}
    (A_{k,m,n,r}, A_{k',m',n',r'})
    =
    c_{\vert m\vert,n,r}^2 
    \,
    \delta(k-k') 
    \,
    \delta_{mm'} \delta_{nn'}
    \delta_{r r'},
\end{align}
where the normalization factors are given by
\begin{align}
    &c^2_{\vert m\vert,n,\pm}
    \equiv
    \frac{1}{2(2n+2\vert m \vert +1)}
    \frac{(n+2\vert m \vert)!\,
    n!}
    {(n+\vert m \vert-1)!\,
    (n+\vert m \vert+1)!}
        \nonumber\\
    & \hspace{0.55\linewidth}
    (m \neq 0, n \geq 0), \\
    &c^2_{0,n,\pm}
    \equiv
    \frac{2}{2n+1}
    \frac{n+1}{n}
    \quad
    (m = 0, n \geq 1),
    \\
    &c^2_{\vert m\vert,n,t}
    \equiv
    \frac{2}
    {2n+2\vert m \vert+1}
    \frac{(n+2\vert m \vert)!}{
     n!}
     \quad
     (m \in \mathbb{Z}, n \geq 0)
    .
\end{align}

We now consider the mode expansion:
\begin{align}
    A^\mu
    =
    \sum_K a_K A^\mu_K,
\end{align}
where the eigenfunctions $A^\mu_{k,m,n,r}$ are written collectively as $A^\mu_K$.
Using that
\begin{align}
    \sqrt{1-z^2}
    \partial_z
    \sqrt{1-z^2}
    Q_m(z)
    =
    -m Q_m(z),
\end{align}
we obtain for the $t$-component of the conserved current, Eq.~\eqref{eq:Jt},
\begin{align}
    J^t(\theta,\phi,t)
    &=
    \int_{-\infty}^\infty dk
    \sum_{m,n,r=\pm} a_{k,m,n,r} 
    \frac{e^{i kt}}{\sqrt{2\pi}}
    \frac{e^{im\phi}}{\sqrt{2\pi}} \nonumber\\
    &~~~~~~~~~~~~~~ \times \frac{-ir}{\sqrt{2}}
    \Big(\frac{1-z}{1+z}\Big)^m
    \partial_z f_{\vert m \vert, n},
\end{align}
where the argument of $f_{\vert m \vert, n}$ involves a sign factor according to Eqs.~\eqref{eq:sol_A_sph} and \eqref{eq:sol_A_sph0}.
Evaluating the expression at $\theta=0$, only the $m=0$ sector remains, leaving
\begin{align}
    J^t(0,\phi,t)
    =
    \int_{-\infty}^\infty dk
    \,
    \frac{e^{i kt}}{\sqrt{2\pi}}
    \sum_{n\geq 1,r=\pm}
    \,
    \frac{i r (n+1)}{\sqrt{4\pi}}
    a_{k,0,n,r}.
\end{align}
The temporal two-point function~\eqref{eq:JJ_t_def} can be calculated to be
\begin{align}
    G_g(t) &=
    \sum_{n\geq 1}
    \frac{n(n+1)(2n+1)}
    {4\pi}
    \int \frac{dk}{2\pi} 
    \frac{e^{ikt}}{k^2 + n(n+1)}
    \nonumber\\
    &=
    \frac{1}{8\pi}
    \sum_{n\geq 1}
    \sqrt{n(n+1)}(2n+1)e^{-\sqrt{n(n+1)}\vert t \vert},
    \label{eq:exact_JJ2}
\end{align}
as quoted in the main text.

\subsection{Analytic checks on \texorpdfstring{$S^2$}{S2}}
\label{sec:further_checks}

We can perform analytic checks for the derived expressions on $S^2$ by using the known facts in two-dimensional field theories. 
From the Ising CFT, the free fermion propagator on $S^2$ is given by (the propagator does not differ by whether the fermion is Dirac or Majorana)
\begin{align}
    \langle \psi(\theta,0) \bar\psi(0,0) \rangle
    =
    \frac{\sigma_1}{4\pi \sin(\theta/2)}.
    \label{eq:psipsi_S2}
\end{align}
On the other hand, from the derived eigenfunctions, we obtain
\begin{align}
    \langle \psi(\theta,0) \bar\psi(0,0) \rangle
    =
    \sigma_1 \cdot \sum_{n\geq 0} \frac{(-1)^n}{\pi\sqrt{2}} \xi_{1/2,n}(-z).
     \label{eq:psipsi_S2_ev}
\end{align}
The two expressions are compared in Fig.~\ref{fig:comp_ferm_2d}, where the $(1,2)$ spinor component is taken.
The summation in Eq.~\eqref{eq:psipsi_S2_ev} is truncated at $n_{\rm max} = 50$ and $200$.
\begin{figure}[htb]
    \centering
    \includegraphics[width=0.8\linewidth]{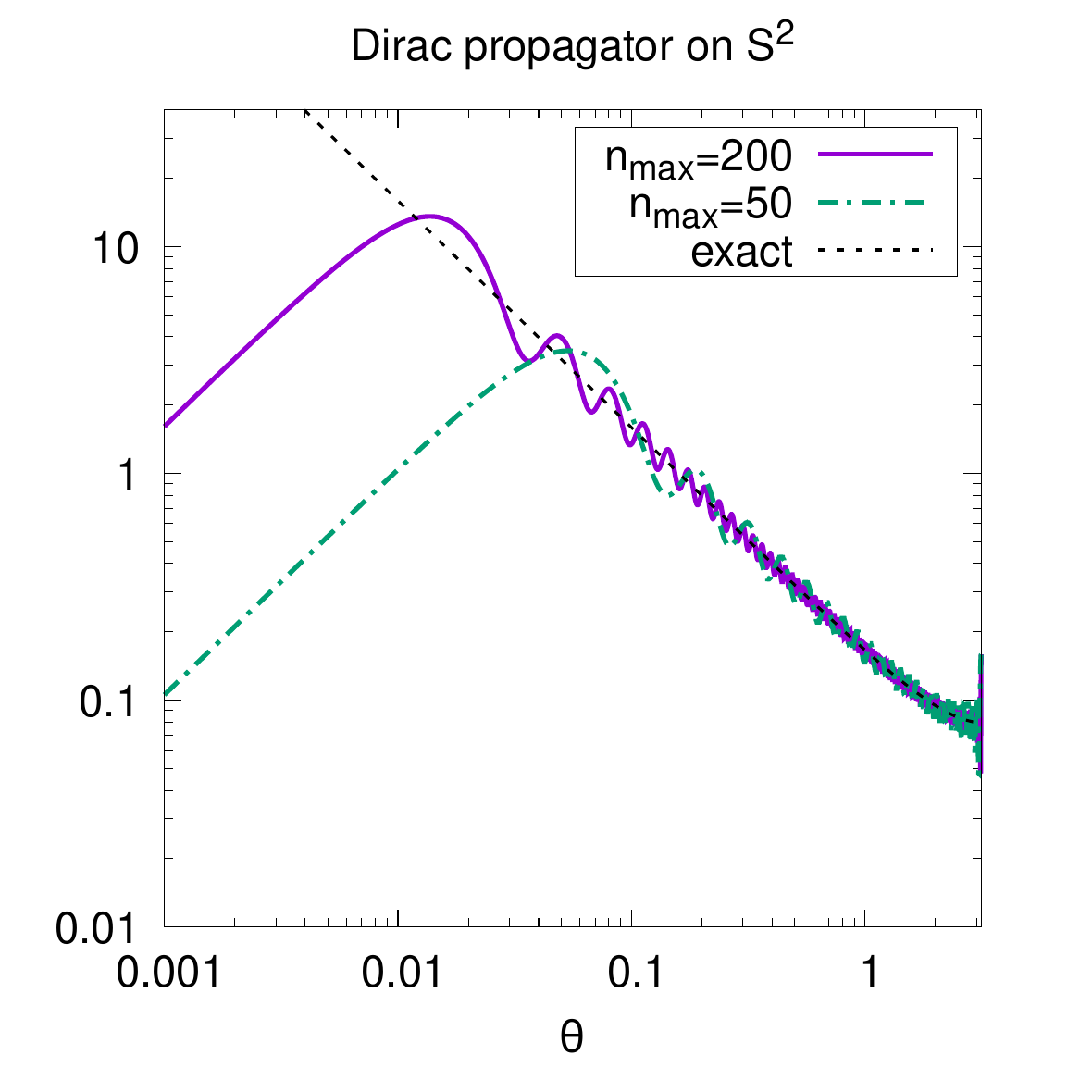}
    \caption{Comparison of the free fermion propagator on $S^2$ between the CFT formula~\eqref{eq:psipsi_S2} and the derived formula~\eqref{eq:psipsi_S2_ev}.
    As we increase the truncation: $1\leq n \leq n_{\rm max}$, we see that the ultraviolet behavior is restored.
    }
    \label{fig:comp_ferm_2d}
\end{figure}
We see that the truncated sum approaches the CFT result as $n_{\rm max} \to \infty$.

As for the pure $U(1)$ gauge theory on $S^2$, the two-point function of $J^t$ gives a delta function because propagation modes do not exist in two dimensions.
For example, on the flat space with volume $V$, one can easily confirm that
\begin{align}
    \frac{1}{g^2}
    \langle J^t(y)J^t(0) \rangle
    =
    \delta^2(y) - \frac{1}{V},
\end{align}
where the constant shift comes from the absence of the constant mode.
On the sphere, again putting the source at $\theta=0$ for simplicity, we obtain
\begin{align}
    \frac{1}{g^2}
    \langle
    J^t(\theta,0)
    J^t(0,0)
    \rangle
    &=
    -\frac{1}{4\pi}
    \sum_{n\geq 1} 
    \frac{2n+1}{n+1} f'_{0,n}(z)
    \nonumber\\
    &=
    \frac{1}{2\pi}
    \Big[
    \delta(z-1)-\frac{1}{2}
    \Big].
    \label{eq:JJ_S2}
\end{align}
The formation of the delta function can be confirmed by integration with test functions, and also graphically as in Fig.~\ref{fig:gauge_delta},
where the summation is truncated at $n_{\rm max}=20, 40$.
\begin{figure}[hbt]
    \centering
    \includegraphics[width=0.8\linewidth]{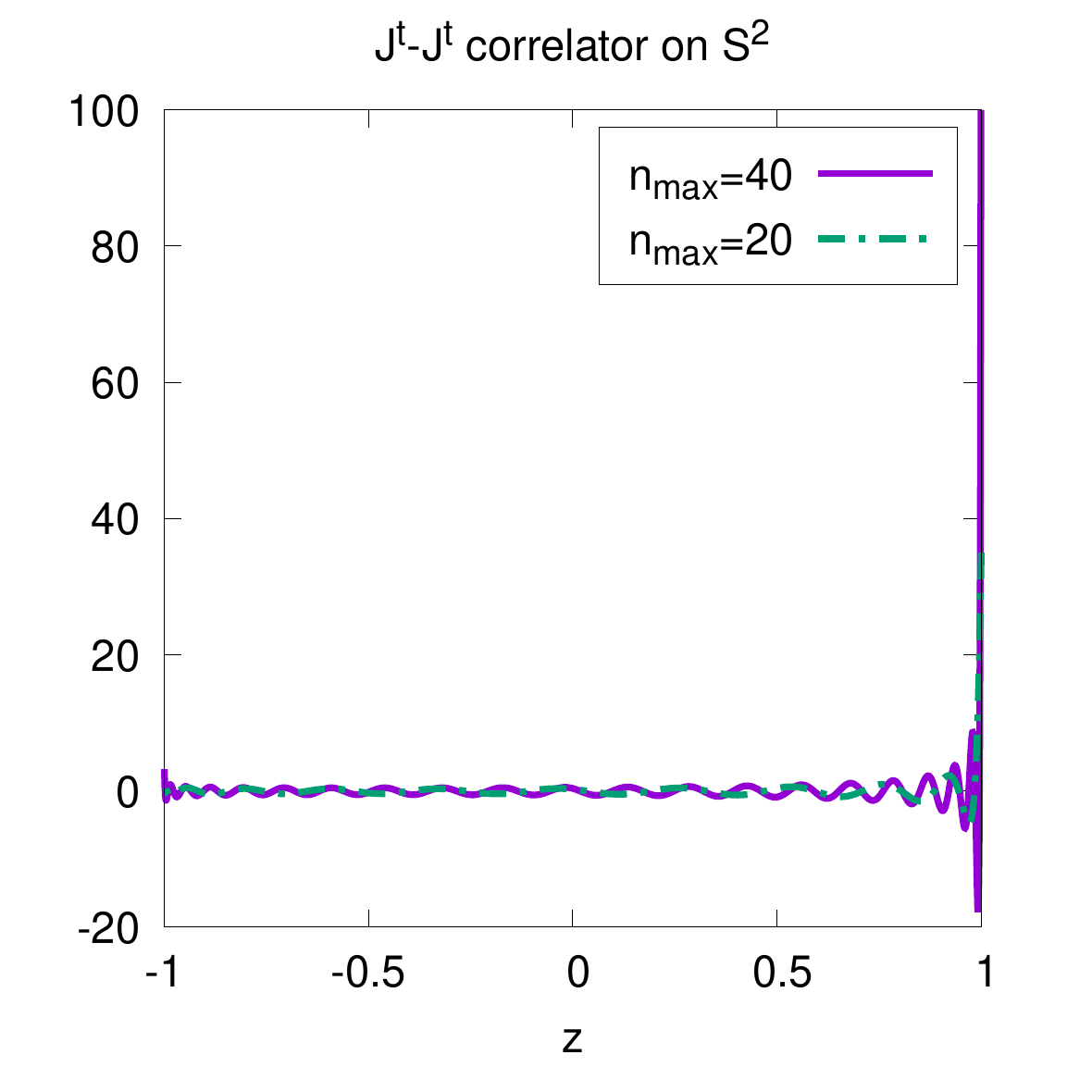}
    \caption{The development of the delta function for the current correlator $(1/g^2)
    \langle
    J^t(\theta,0)
    J^t(0,0)
    \rangle$ on $S^2$.
    The truncation in the summation is varied as $n_{\rm max} = 20, 40$ in Eq.~\eqref{eq:JJ_S2}.
    }
    \label{fig:gauge_delta}
\end{figure}

\bibliography{ref}

\end{document}